\documentclass[journal,twocolumn,10pt,twoside]{IEEEtran}
\normalsize

\ifCLASSINFOpdf
\else
\fi

\usepackage{amsmath}
\usepackage{amssymb}
\usepackage{cite}
\usepackage{graphicx}
\usepackage{algorithm}
\usepackage{algpseudocode}
\usepackage{subfigure}
\usepackage{multirow}
\usepackage{longtable}
\usepackage{threeparttable}
\usepackage{caption3}
\usepackage{array}
\usepackage{cases}
\usepackage{color}
\usepackage[T1]{fontenc}
\usepackage[utf8]{inputenc}
\usepackage{authblk}
\usepackage{subfigmat}
\usepackage{fixltx2e}

\newtheorem{remark}{Remark}

\newtheorem{observation}{Observation}

\newcommand{\Tr}{\mathrm{Tr}}
\newcommand{\ID}{\mathrm{ID}}
\newcommand{\EH}{\mathrm{EH}}
\newcommand{\rf}{\mathrm{rf}}
\newcommand{\dc}{\mathrm{dc}}
\newcommand{\In}{\mathrm{in}}
\newcommand{\out}{\mathrm{out}}
\newcommand{\ant}{\mathrm{ant}}
\newcommand{\R}{\textnormal{R}}
\newcommand{\E}{\textnormal{E}}
\newcommand{\sinc}{\mathrm{sinc}}
\newcommand\mycom[2]{\genfrac{}{}{0pt}{}{#1}{#2}}

\begin{document}
\title{Fundamentals of Wireless Information and Power Transfer: From RF Energy Harvester Models to Signal and System Designs}
\author{Bruno Clerckx, \textit{Senior Member, IEEE}, Rui Zhang, \textit{Fellow, IEEE}, Robert Schober, \textit{Fellow, IEEE}, Derrick Wing Kwan Ng, \textit{Senior Member, IEEE}, Dong In Kim, \textit{Senior Member, IEEE}, and H. Vincent Poor, \textit{Fellow, IEEE}
\thanks{B. Clerckx is with the Electrical and Electronic Engineering Department at Imperial College London, London SW7 2AZ, UK (email: b.clerckx@imperial.ac.uk).
\par R. Zhang is the Department of Electrical and Computer Engineering,
National University of Singapore, Singapore 117583 (e-mail:
elezhang@nus.edu.sg).
\par R. Schober is with the Institute of Digital
Communications, Friedrich-Alexander-University Erlangen-Nurnberg
(FAU), Germany (email: robert.schober@fau.de).
\par D. W. K. Ng is with the School of Electrical Engineering and Telecommunications,
University of New South Wales, Sydney, NSW 2052, Australia
(email: w.k.ng@unsw.edu.au).
\par D. I. Kim is with the School of Information and Communication Engineering,
Sungkyunkwan University (SKKU), Korea (email: dikim@skku.ac.kr).
\par H. V. Poor is with the Department of Electrical Engineering, Princeton
University, Princeton, NJ 08544 USA (e-mail: poor@princeton.edu).
\par This work has been partially supported by the EPSRC of UK, under grant EP/P003885/1. }}

\maketitle

\begin{abstract}
Radio waves carry both energy and information simultaneously. Nevertheless, Radio-Frequency (RF) transmission of these quantities have traditionally been treated separately. Currently, we are experiencing a paradigm shift in wireless network design, namely unifying wireless transmission of information and power so as to make the best use of the RF spectrum and radiations as well as the network infrastructure for the dual purpose of communicating and energizing. In this paper, we review and discuss recent progress on laying the foundations of the envisioned dual purpose networks by establishing a signal theory and design for Wireless Information and Power Transmission (WIPT) and identifying the fundamental tradeoff between conveying information and power wirelessly. We start with an overview of WIPT challenges and technologies, namely Simultaneous Wireless Information and Power Transfer (SWIPT), Wirelessly Powered Communication Network (WPCN), and Wirelessly Powered Backscatter Communication (WPBC). We then characterize energy harvesters and show how WIPT signal and system designs crucially revolve around the underlying energy harvester model. To that end, we highlight three different energy harvester models, namely one linear model and two nonlinear models, and show how WIPT designs differ for each of them in single-user and multi-user deployments. Topics discussed include rate-energy region characterization, transmitter and receiver architecture, waveform design, modulation, beamforming and input distribution optimizations, resource allocation, and RF spectrum use.  We discuss and check the validity of the different energy harvester models and the resulting signal theory and design based on circuit simulations, prototyping and experimentation. We also point out numerous directions that are promising for future research.
\end{abstract}


\IEEEpeerreviewmaketitle

\section{Introduction}\label{Intro_section}

\par Wireless communications via Radio-Frequency (RF) radiation has been around for more than a century and has significantly shaped our society in the past 40 years. Wireless is however not limited to communications. Wireless powering of devices using near-field Inductive Power Transfer has become a reality with several commercially available products and standards. However, its range is severely limited (less than one meter). On the other hand, far-field Wireless Power Transfer (WPT) via RF (as in wireless communication) could be used over longer ranges. It has long been regarded as a possibility for energising low-power devices but it is only recently that it has become recognized as feasible due to reductions in the power requirements of electronics and smart devices \cite{Smith:2013,Hemour:2014}. Indeed, in 20 years from now, according to Koomey's law \cite{Koomey:2011}, the amount of energy needed for a given computing task will fall by a factor of 10000 compared to what it is now, thus further continuing the trend towards low-power devices. Moreover, the world will see the emergence of trillions of Internet-of-Things (IoT) devices. This explosion of low-power devices calls for a re-thinking of wireless network design.
\par Recent research advocates that the future of wireless networking goes beyond conventional communication-centric transmission. In the same way as wireless (via RF) has disrupted mobile communications for the last 40 years, wireless (via RF) will disrupt the delivery of mobile power. However, current wireless networks have been designed for communication purposes only. While mobile communication has become a relatively mature technology, currently evolving towards its fifth generation, the development of mobile power is in its infancy and has not even reached its first generation. Today, not a single standard on far-field WPT exists.
Wireless power will bring numerous new opportunities: no wires, no contacts, no batteries, genuine mobility and a perpetual, predictable, dedicated, on-demand, and reliable energy supply as opposed to intermittent ambient energy-harvesting technologies (e.g. solar, thermal, vibration). This is highly relevant in future networks with ubiquitous and autonomous low-power and energy limited devices, device-to-device communications, and the IoT with massive connections.
\par Interestingly, although radio waves carry both energy and information simultaneously, RF transmission of these quantities have traditionally been treated separately. Imagine instead a wireless network, e.g. WiFi, in which information and energy flow together through the wireless medium. Wireless communication, or Wireless Information Transfer (WIT), and WPT would then refer to two extreme strategies, respectively, targeting communication-only and power-only. A unified design of Wireless Information and Power Transmission (WIPT) would on the other hand have the ability to softly evolve and compromise in between those two extremes to make the best use of the RF spectrum/radiation and the network infrastructure to communicate and energize. This will enable trillions of low-power devices to be connected and powered anywhere, anytime, and on the move.
\par The integration of wireless power and wireless communications brings new challenges and opportunities, and calls for a paradigm shift in wireless network design. As a result, numerous new research problems need to be addressed that cover a wide range of disciplines including communication theory, information theory, circuit theory, RF design, signal processing, protocol design, optimization, prototyping, and experimentation.

\subsection{Overview of WIPT Challenges and Technologies}

\par WIT and WPT are fundamental building blocks of WIPT and the design of efficient WIPT networks fundamentally relies on the ability to design efficient WIT and WPT. In the last 40 years, WIT has seen significant advances in RF theory and signal theory. Traditional research on WPT in the last few decades has focused extensively on RF theories and techniques concerning the energy receiver with the design of efficient RF, circuit, antenna, rectifier, and power management unit solutions \cite{OptBehaviour,Valenta:2014,Costanzo:2016}, but recently a new and complementary line of research on communications and signal design for WPT has emerged in the communication literature \cite{Zeng:2017}. Moreover, there has been a growing interest in bridging RF, signal, and system designs in order to bring those two communities closer together and to get a better understanding of the fundamental building blocks of an efficient WPT network architecture \cite{Clerckx:2018}.

\par The engineering requirements and design challenges of the envisioned network are numerous: 1) \textit{Range}: Delivery of wireless power at distances of 5-100 meters (m) for indoor/outdoor charging of low-power devices; 2) \textit{Efficiency}: Boosting the end-to-end power transfer efficiency (up to a fraction of a percent/a few percent), or equivalently the DC power level at the output of the rectenna(s) for a given transmit power; 3) \textit{Non-line of sight (NLoS)}: Support of Line of sight (LoS) and NLoS to widen the practical applications of WIPT networks; 4) \textit{Mobility support}: Support of mobile receivers, at least for those at pedestrian speed; 5) \textit{Ubiquitous accessibility}: Support of ubiquitous power accessibility within the network coverage area;  6) \textit{Safety and health}: Resolving the safety and health issues of RF systems and compliance with the regulations; 7) \textit{Energy consumption}: Limitation of the energy consumption of energy-constrained RF powered devices; 8) \textit{Seamless integration of wireless communication and wireless power}: Interoperability between wireless communication and wireless power via a unified WIPT.

\par Solutions to tackle challenges 1)-7) are being researched and have been discussed extensively in \cite{Shinohara:2014,Carvalho:2014,Costanzo:2016,Zeng:2017,Clerckx:2018}. They cover a wide range of areas spanning sensors, devices, RF, communication, signal and system designs for WPT. This survey article targets challenge 8) by reviewing the fundamentals of WIPT signal and system designs. In WPT and WIT, the emphasis of the system design is to exclusively deliver energy and information, respectively. On the contrary, in WIPT, both energy and information are to be delivered. A WIPT system should therefore be designed such that the RF radiation and the RF spectrum are exploited in the most efficient manner to deliver both information and energy. Such a system design requires the characterization of the fundamental tradeoff between how much information and how much energy can be delivered in a wireless network and how signals should be designed to achieve the best possible tradeoff between them.
\par
\begin{figure*}
   \centerline{\includegraphics[width=14cm]{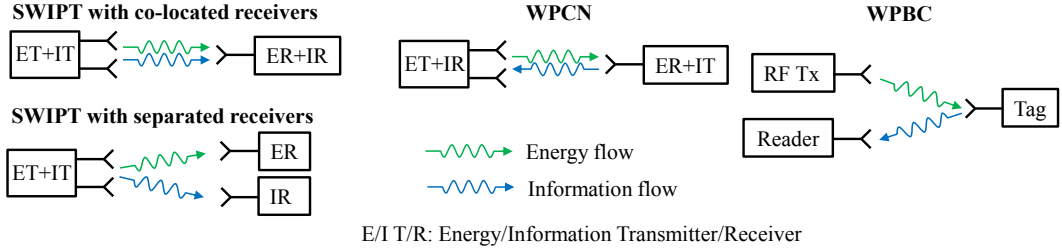}}
  \caption{Different WIPT architectures.}
  \label{SWIPT_figure}
\end{figure*}

 As illustrated in Fig. \ref{SWIPT_figure}, WIPT can be categorized into three different types:
\begin{itemize}
\item \textit{Simultaneous Wireless Information and Power Transfer (SWIPT)}: Energy and information are simultaneously transferred in the downlink from one or multiple access points to one or multiple receivers. The Energy Receiver(s) (ER) and Information Receiver(s) (IR) can be co-located or separated. In SWIPT with separated receivers, ER and IR are different devices, the former being a low-power device being charged, the latter being a device receiving data. In SWIPT with co-located receivers, each receiver is a single low-power device that is simultaneously being charged and receiving data.
\item \textit{Wirelessly Powered Communication Network (WPCN)}: Energy is transferred in the downlink and information is transferred in the uplink. The receiver is a low-power device that harvests energy in the downlink and uses it to send data in the uplink.
\item \textit{Wirelessly Powered Backscatter Communication (WPBC)}: Energy is transferred in the downlink and information is transferred in the uplink but backscatter modulation at a tag is used to reflect and modulate the incoming RF signal for communication with a reader. Since tags do not require oscillators to generate carrier signals, backscatter communications benefit from orders-of-magnitude lower power consumption than conventional radio communications.
\end{itemize}
Moreover, a network could also include a mixture of the above three types of transmissions with multiple co-located and/or separated Energy Transmitter(s) (ET) and Information Transmitter(s) (IT).

\subsection{Objectives and Organization}

\par This paper reviews and summarizes recent advances and contributions in the area of WIPT. The main objective of this article is to give a systematic treatment of signal theory and design for WIPT and use it to characterize the fundamental tradeoff between conveying information and energy in a wireless network. This tradeoff is commonly referred to as rate-energy (R-E) tradeoff. Various review papers on WIPT have appeared in past years \cite{Bi:2015,Tabassum:2015,KHuang:2015,Bi:2016,KHuang:2016,Krikidis:2014,Ding:2015,Chen:2015,Lu:2015,Ulukus:2015,Niyato:2017}. Emphasis was put at that time on characterizing the R-E tradeoff under the assumption of a very simple \textit{linear model} of the energy harvester. Interestingly, the importance of the energy harvester model for WIPT design was never raised and the validity of this linear model never questioned in that WIPT literature. In recent years, there has been an increasing interest in the WIPT literature to depart from the linear model. However what we know about WIPT design from those review papers is fundamentally rooted in the underlying linear model. It turns out that WIPT design radically changes once we change the energy harvester model and adopt more realistic \textit{nonlinear models} of the energy harvester.
\par Hence, in contrast to those existing tutorial and review papers, we here aim at showing how crucial the energy harvester model is to WIPT signal and system designs and how WIPT signal and system designs revolve around the underlying energy harvester model. To that end, we highlight three different energy harvester models, namely one \textit{linear model} and two \textit{nonlinear models}, and show how WIPT designs differ for each of them. In particular, we show how the modeling of the energy harvester can have tremendous influence on the design of the Physical (PHY) and Medium Access Control (MAC) layers of WIPT networks. We rigorously review how the different models can favor different waveforms, modulations, input distributions, beamforming, transceiver architectures, and resource allocation strategies as well as a different use of the RF spectrum. We first consider single-user (point-to-point) WIPT and then extend to multi-user scenarios. We discuss the validity of the different energy harvester models and the resulting signal and system designs through experimentation and prototyping. Finally, we point out directions that are promising for future research.

\par The rest of this article is organized as follows. In the next subsection, we first give some insights into the crucial role of energy harvester modeling and its impact onto signal designs. We then jump into the core parts of the paper. Section \ref{Rectenna_section} introduces three models for the energy harvester (rectenna), namely the diode linear model, the diode nonlinear model, and the saturation nonlinear model. Section \ref{single_user_WIPT_section} is dedicated to the study of the fundamental tradeoff between rate and energy in single-user (point-to-point) WIPT for each of the three rectenna models. Special emphasis is given to how deeply the rectenna model influences the R-E tradeoff and WIPT signal and system design. Section \ref{multi_user_WIPT_section} extends the discussion to multi-user WIPT. Section \ref{proto_section} discusses recent prototyping and experimentation efforts to validate the signal theory and designs. Section \ref{conclusions} concludes the paper.

\par Throughout the paper, a special emphasis is put on SWIPT as it can be seen as the most involved and disruptive scenario, where wireless communications and wireless power are closely intertwined. Nevertheless, the analysis and ideas reviewed in the paper can also find applications in WPCN and WPBC, as pointed out throughout the manuscript.

\subsection{The Crucial Role of Energy Harvester Modeling}

\begin{figure}
   \centerline{\includegraphics[width=0.9\columnwidth]{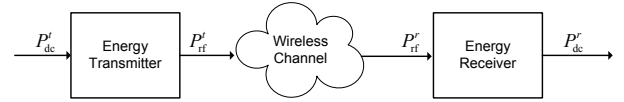}}
	\caption{The block diagram of a generic WPT system \cite{Zeng:2017}.}\label{F:BasicArchitecture}
\end{figure}

\par In order to motivate the importance of the energy harvester modeling, recall first the block diagram of a generic WPT system illustrated in Fig. \ref{F:BasicArchitecture}. The end-to-end power transfer efficiency $e$ can be decomposed as
\begin{align}\label{e_equation}
e=\frac{P_{\dc}^r}{P_{\dc}^t}=\underbrace{\frac{P_{\rf}^t}{P_{\dc}^t}}_{e_1}\underbrace{\frac{P_{\rf}^r}{P_{\rf}^t}}_{e_2}\underbrace{\frac{P_{\dc}^r}{P_{\rf}^r}}_{e_3},
\end{align}
where $e_1$, $e_2$, and $e_3$ denote the DC-to-RF, RF-to-RF, and RF-to-DC power conversion/transmission efficiency, respectively.

\par A natural approach to come up with an efficient WPT architecture would be to concatenate techniques designed specifically to maximize $e_1$, $e_2$, and $e_3$. One could therefore use an efficient Power Amplifier (PA), smart channel-adaptive signals, and an efficient rectenna to maximize $e_1$, $e_2$, and $e_3$, respectively. Doing so, the RF and signal designs are completely decoupled. WPT/WIPT RF designers would deal with efficient PA and rectenna designs and WPT/WIPT signal designers focus on maximizing $e_2$ assuming $e_1$ and $e_3$ constant, i.e., assuming $e_1$ and $e_3$ are not a function of the transmit/received signals but only a function of the PA and rectenna designs, respectively. Though not explicitly stated this way, this is the design philosophy adopted in the early works on SWIPT, WPCN and WPBC, see e.g. \cite{Varshney:2008,Grover:2010,Zhang:2013,Son:2014,Xu:2014,Zhou:2013,Liu:2013,Park:2013,Park:2014,Park:2015,Huang:2013,Zhou:2014,Ng:2013,Nasir:2013,Huang:2015,Ju:2014,Lee:2016,Boyer:2014,Yang:2015,Bi:2015,Tabassum:2015,KHuang:2015,Bi:2016,Krikidis:2014,Ding:2015,Chen:2015,Lu:2015}.
\par SWIPT was first considered in \cite{Varshney:2008}. The tradeoff between information rate and delivered energy, the so-called R-E region, was characterized for point-to-point discrete channels, and a Gaussian channel subject to an amplitude constraint on the input. SWIPT was then studied in a frequency-selective Gaussian channel under an average power constraint in \cite{Grover:2010}. In \cite{Zhang:2013}, the term SWIPT was first coined and SWIPT was investigated for multi-user MIMO systems, where practical receivers to realize both RF energy harvesting and information decoding were proposed. Since then, SWIPT has attracted significant interests in the communication literature with works covering a wide range of topics, among others MIMO broadcasting \cite{Son:2014,Xu:2014}, architecture \cite{Zhou:2013,Liu:2013}, interference channel \cite{Park:2013,Park:2014,Park:2015}, broadband system \cite{Huang:2013,Zhou:2014,Ng:2013}, relaying \cite{Nasir:2013,Huang:2015}. In parallel, much attention has been drawn to WPCN \cite{Ju:2014,Lee:2016} and WPBC \cite{Boyer:2014,Yang:2015}.
\par Interestingly, while the above literature addresses complicated scenarios with multiple transmitters and receivers and complicated R-E tradeoff characterizations, results are based on the assumptions that $e_1$ and $e_3$ are constant. Indeed, the DC-to-RF conversion efficiency $e_1$ has been assumed equal to unity and the energy harvester has been abstracted using a linear relationship stating that the output DC power of the energy harvester is equal to its input RF power multiplied by a constant RF-to-DC conversion efficiency $e_3$ \cite{Zeng:2017}. Such a \textit{linear model} for the energy harvester has the benefit of being analytically easily tractable.
\par Another approach to designing efficient WPT and WIPT architectures has emerged more recently and relies on observations made in the RF literature that the RF-to-DC conversion efficiency $e_3$ is not a constant but a \textit{nonlinear} function of the input signal (power and shape) \cite{Trotter:2009,Trotter:2010,Boaventura:2011,Collado:2014,Valenta:2015,Clerckx:2018}. Assuming $e_3$ constant is indeed over-simplified and is not validated by circuit simulations and measurements. This observation has as consequence that the maximization of $e$ is not achieved by maximizing $e_1$, $e_2$ and $e_3$ independently from each other, and therefore, simply concatenating an efficient PA, an $e_2$-maximizing signal, and an efficient rectenna \cite{Clerckx:2018}. Efficiencies $e_1$, $e_2$ and $e_3$ are indeed coupled with each other due to the energy harvester nonlinearity \cite{Clerckx:2016b,Boshkovska:2015,Zeng:2017}. The RF-to-DC conversion efficiency $e_3$ is not only a function of the rectenna design but also of its input signal shape and power and therefore a function of the transmit signal (beamformer, waveform, modulation, power allocation) and the wireless channel state. Similarly, $e_2$ depends on the transmit signal and the channel state and so does $e_1$, since it is a function of the transmit signal Peak-to-Average Power Ratio (PAPR). Hence, signal design not only influences $e_2$ but also $e_1$ and $e_3$ in general settings. Being able to predict the influence of the signal design on $e_1$ and $e_3$ requires the development of \textit{nonlinear models} for the PA and the energy harvester, respectively. Of particular interest in this paper is the modeling of the energy harvester and the influence of the signal design on $e_2$ and $e_3$.

\emph{Notations:} In this paper, scalars are denoted by italic letters. Boldface lower- and upper-case letters denote vectors and matrices, respectively. $\mathbb{C}^{M\times N}$ denotes the space of $M\times N$ complex matrices. $j$ denotes the imaginary unit, i.e., $j^2=-1$. $\mathbb{E}[\cdot]$ denotes statistical expectation and $\Re\{\cdot\}$ represents the real part of a complex number. $\mathbf{I}_M$ denotes an $M\times M$ identity matrix and $\mathbf{0}$ denotes an all-zero vector/matrix. For an arbitrary-size matrix $\mathbf{A}$,  its complex conjugate, transpose, Hermitian transpose, and Frobenius  norm are respectively denoted as $\mathbf A^*$, $\mathbf{A}^{T}$, $\mathbf{A}^{H}$, and $\|\mathbf{A}\|_F$. $[\mathbf A]_{im}$ denotes the $(i,m)$th element of matrix $\mathbf A$. For a square Hermitian matrix $\mathbf{S}$, $\mathrm{Tr}(\mathbf{S})$ denotes its trace, while $\lambda_{\max}(\mathbf S)$ and $\mathbf v_{\max}(\mathbf S)$ denote its largest eigenvalue and the corresponding eigenvector, respectively. In the context of random variables, i.i.d. stands for independent and identically distributed. The distribution of a Circularly Symmetric Complex Gaussian (CSCG) random variable with zero-mean and variance $\sigma^2$ is denoted by $\mathcal{CN}(0,\sigma^2)$; hence with the real/imaginary part distributed as $\mathcal{N}(0,\sigma^2/2)$. $\sim$ stands for ``distributed as''. We use the notation $\textnormal{sinc}\left(t\right)=\frac{\sin\left(\pi t\right)}{\pi t}$.

\section{Analytical Models for the Rectenna}\label{Rectenna_section}

\par The energy receiver in Fig. \ref{F:BasicArchitecture} consists of an energy harvester comprising a rectenna (antenna and rectifier) and a power
management unit (PMU). Since the quasi-totality of electronics requires a DC power source, a rectifier is required to convert RF to DC. The recovered DC power then either powers a low-power device directly, or is stored in a battery or a super-capacitor for higher power low duty-cycle operations. It can also be managed by a DC-to-DC converter as part of the PMU before being stored. In the sequel, we will not discuss the PMU but only the rectenna models. We first start by giving a short overview of rectennas before jumping into the rectenna models.

\subsection{Rectenna Behavior}\label{antenna_model_subsection}

\begin{figure}
\centerline{\includegraphics[width=0.9\columnwidth]{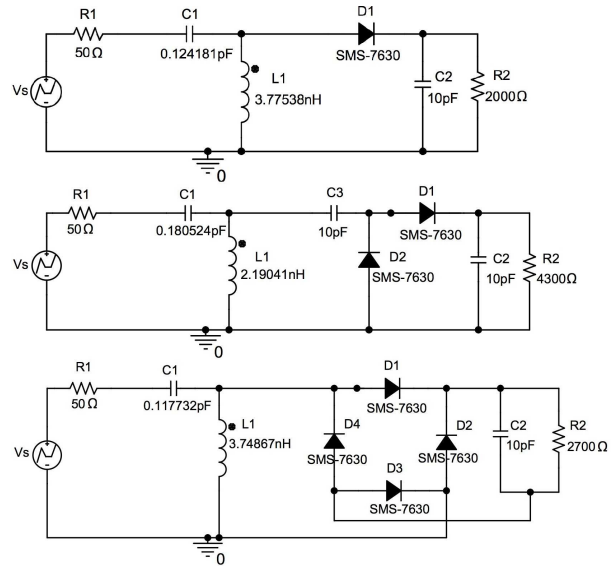}}
  \caption{Examples of single series, voltage doubler, and diode bridge rectifiers, designed for an average RF input power of -20dBm at 5.18GHz. $v_{\mathrm{s}}$ is the voltage source of the antenna \cite{Clerckx:2017}. R1 models the antenna impedance. C1 and L1 form the matching network. D1, D2, D3, and D4 refer to the Schottky diodes. C2 and R2 form the low-pass filter with R2 being the output load.}
  \label{rectenna_circuit}
\end{figure}

\par A rectenna harvests electromagnetic energy, then rectifies and filters it using a low pass filter. Various rectifier technologies (including the popular Schottky diodes, CMOS, active rectification, spindiode, backward tunnel diodes) and topologies (with single and multiple diode rectifier) exist \cite{OptBehaviour,Valenta:2014,Costanzo:2016}. Examples of single series, voltage doubler and diode bridge rectifiers consisting of 1, 2 and 4 Schottky diodes respectively are illustrated in Fig. \ref{rectenna_circuit} \cite{Clerckx:2017}. In its simplest form, the single series rectifier is made of a matching network (to match the antenna impedance to the rectifier input impedance) followed by a single diode and a low-pass filter, as illustrated by the circuit at the top in Fig. \ref{rectenna_circuit}.

\par Assuming $P_{\rf}^t$=1 Watt (W), 5-dBi Tx/Rx antenna gain, a continuous wave (CW) at 915MHz, $e_3$ of state-of-the-art rectifiers is 50\% at 1m, 25\% at 10m and about 5\% at 30m \cite{Hemour:2014}. Hence, $e_3$ (and $e_2$ as well) decreases as the range increases. Viewed differently, this implies that $e_3$ decreases as the input power $P_{\rf}^{r}$ to the rectifier decreases. Indeed, $e_3$ of state-of-the-art rectifiers drops from 80\% at $P_{\rf}^{r}=$10 mW to 40\% at 100 $\mu$W, 20\% at 10 $\mu$W and 2\% at 1 $\mu$W \cite{Valenta:2014, Hemour:2014}. This is due to the rectifier sensitivity with the diode not being easily turned on at low input power. For typical input powers between 1 $\mu$W and 1 mW, low barrier Schottky diodes remain the most competitive and popular technology \cite{Valenta:2014,Costanzo:2016}. A single diode is commonly preferred at low power (1-500 $\mu$W) because the amount of input power required to switch on the rectifier is minimized. Multiple diodes (voltage doubler/diode bridge/charge pump) are on the other hand favoured at higher input power, typically above 500$\mu$W \cite{Costanzo:2016,OptBehaviour}. Topologies using multiple rectifying devices each one optimized for a different range of input power levels also exist and can enlarge the operating range versus input power variations \cite{Sun:2013}. This can be achieved using e.g. a single-diode rectifier at low input power and multiple diodes rectifier at higher power.

\begin{figure}
\centerline{\includegraphics[width=0.9\columnwidth]{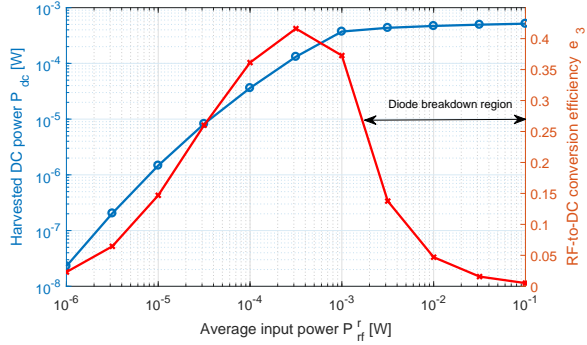}}
  \caption{Harvested DC power $P_{\dc}^r$ vs average input power $P_{\rf}^{r}$ and RF-to-DC conversion efficiency $e_3$ with a single-series rectifier obtained from circuit simulations \cite{Clerckx:2018b}. The input signal is a continuous wave at 5.18 GHz and rectifier designed for -20dBm input power.}
  \label{Pdc_Pin}
\end{figure}
\par Fig. \ref{Pdc_Pin} illustrates the dependency of $e_3$ to the average signal power at the input of the rectifier. Using circuit simulations and a single-series rectifier similar to the one at the top of Fig. \ref{rectenna_circuit}, we plot the DC power $P_{\dc}^{r}$ harvested at the load as a function of the input power $P_{\rf}^r$ to the rectifier when a CW (i.e.\ a single sinewave) signal is used for excitation \cite{Clerckx:2018b}. We also display the RF-to-DC conversion efficiency $e_3=P_{\dc}^{r}/P_{\rf}^{r}$. This circuit was designed for 10$\mu$W input power but as we can see it can operate typically between 1$\mu$W and 1mW. Clearly, the RF-to-DC conversion efficiency $e_3$ is not a constant, but depends on the input power level. It is about 2\% at 1$\mu$W, 15\% at 10$\mu$W and 35\% at 100$\mu$W, which is inline with the values reported from the literature in the previous paragraph. Beyond 1mW input power, the output DC power saturates and $e_3$ suddenly significantly drops, i.e., the rectifier enters the diode breakdown region. Indeed, the diode SMS-7630 becomes reverse biased at 2 Volts (V), corresponding to an input power of about 1mW. To operate beyond 1mW, a rectifier with multiple diodes (similarly to the ones in Fig. \ref{rectenna_circuit}) would be preferred so as to avoid the saturation problem \cite{OptBehaviour,Costanzo:2016,Sun:2013}.

\par
\begin{figure}
\centerline{\includegraphics[width=0.9\columnwidth]{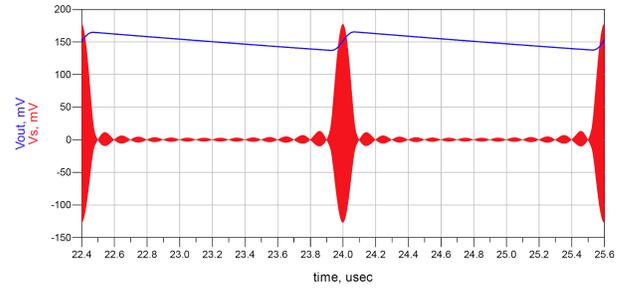}}
  \caption{Example of the multisine waveform (red) used as a voltage source $v_{\mathrm{s}}$ at the input of the rectenna and the rectified output voltage $v_{\mathrm{out}}$ (blue). See also Fig. \ref{antenna_model} for an illustration of a rectenna and the corresponding $v_{\mathrm{s}}$ and $v_{\mathrm{out}}$.}
  \label{pow}
\end{figure}

 The above discussion illustrates the dependency of $e_3$ on the rectifier design and the average received signal power level $P_{\rf}^r$. Actually $e_3$ is also a function of the rectifier's input signal shape and not only power. This was first highlighted in \cite{Trotter:2009,Trotter:2010}, wherein the authors proposed the use of a multisine waveform instead of a continuous wave (single sinewave) to provide a higher charge pump efficiency and thus to increase the range of RFID readers. A multisine is characterized by a high PAPR, and the envelope of the transmitted RF signal is designed so that there are large peaks, while the average power is kept the same as in the continuous wave case. Consider indeed multiple in-phase sinewaves (with equal magnitudes) at frequencies $f_n=f_0+n\Delta_f$, $n=0,\ldots,N-1$, as the voltage source of the rectenna. As the number of tones $N$ increases, the time domain waveform appears as a sequence of pulses with a period equal to $1/\Delta_f$ illustrated by the red curve in Fig. \ref{pow}. The signal power is therefore concentrated into a series of high energy pulses, each of which triggers the diode that then conducts and helps charging the output capacitor. Once a pulse has passed, the diode stops conducting and the capacitor is discharging, as illustrated by the blue curve in Fig. \ref{pow}. The larger the number of tones $N$, the larger is the magnitude of the pulses and therefore the larger is the output voltage at the time of discharge. Since peaks of high power drive the rectenna with a much higher efficiency than the average low level input, they contribute more to the output DC voltage, and the rectifier sensitivity, range and RF-to-DC conversion efficiency $e_3$ increase. A more systematic way to design and optimize multisine waveforms for WPT was proposed in \cite{Clerckx:2016b}. Though limited to deterministic multisine signals, the discussion illustrates a key starting point of the paper, namely the fact that the RF-to-DC conversion efficiency $e_3$ is influenced by the input signal shape and power to the rectifier.

\par Modeling the dependency of $e_3$ on the input signal shape and power is very challenging. This is so because RF-based energy harvesting circuits consist of various components such as resistors, capacitors, and diodes that introduce various nonlinearities \cite{Valenta:2014,Costanzo:2016,CN:EH_measurement_2,JR:EH_measurement_1}. This ultimately makes rectenna modeling and analysis an important and challenging research area \cite{Costanzo:2016,CN:EH_measurement_2,JR:EH_measurement_1}. Moreover the practical implementation of rectenna is hard and subject to several losses due to threshold and reverse-breakdown voltages, devices parasitics, impedance matching, and harmonic generation \cite{Valenta:2014}. In the sequel, we introduce various models for the rectenna. The first two models, the so-called \textit{diode linear model} and \textit{diode nonlinear model}, are driven by the physics of the diode and relate the output DC current/power to the input signal through the diode current-voltage (I-V) characteristics \cite{Clerckx:2016b}. The diode linear model is a particular case of the diode nonlinear model and is obtained by ignoring the diode nonlinearity \cite{Zhou:2013}. The third model, the so-called \textit{saturation nonlinear model}, models the saturation of the output DC power at large RF input power due to the diode breakdown. In contrast to the first two models, the third model is circuit-specific and obtained via curve fitting based on measured data \cite{Boshkovska:2015}.

\par It is important to note that more complicated models can be found in the RF literature, where one could for instance derive mathematical (differential) equations to describe the exact input-output characteristic of an RF-based energy harvesting circuit based on its schematic such as in Fig. \ref{rectenna_circuit}.  However, RF-based energy harvesting circuits may consist of various multistage rectifying circuits. This leads to complicated analytical expressions which are intractable for signal and resource allocation algorithm design. More importantly, such an approach may rely on specific implementation details of energy harvesting circuits and the corresponding mathematical expressions may differ significantly across different types of energy harvesting circuits. In contrast, the three models described in the sequel are driven by a tradeoff between accuracy and tractability. They may appear oversimplified from an RF perspective but the goal here is to extract the key elements of the energy harvester that influences signal and resource allocation design and enables insights for signal and system designs.

\subsection{The Antenna Model}\label{antenna_model_subsection}

\begin{figure}
   \centerline{\includegraphics[width=0.9\columnwidth]{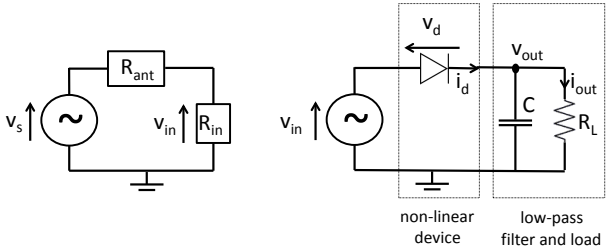}}
  \caption{Antenna equivalent circuit (left) and a single diode rectifier (right) \cite{Clerckx:2016b}. The rectifier comprises a non-linear device (diode) and a low-pass filter (consisting of a capacitor C and a load R\textsubscript{L}).}
  \label{antenna_model}
\end{figure}

A lossless antenna is modeled as a voltage source $v_{\mathrm{s}}(t)$ followed by a series resistance\footnote{Assumed real for simplicity. A more general model can be found in \cite{Clerckx:2017}.} $R_{\ant}$ (Fig. \ref{antenna_model} left hand side). Let $Z_{\In}\!=\! R_{\In} + j X_{\In}$ denote the input impedance of the rectifier and the matching network. Let $y_{\rf}(t)$ also denote the RF signal impinging on the receive antenna. Assuming perfect matching ($R_{\In} = R_{\ant}$, $X_{\In} = 0$), the available RF power $P_{\rf}^r$ is transferred to the rectifier and absorbed by $R_{\In}$, so that $P_{\rf}^r = \mathbb{E}\big[y_{\rf}(t)^2\big]\!=\!\mathbb{E}\big[v_{\In}(t)^2\big]/R_{\In}$, $v_{\In}(t)\!=\!v_{\mathrm{s}}(t)/2$, and $v_{\In}(t)\!=\!y_{\rf}(t)\sqrt{ R_{\In}}\!=\!y_{\rf}(t)\sqrt{ R_{\ant}}$. We also assume that the antenna noise is too small to be harvested.

\subsection{The Diode Linear and Nonlinear Models}\label{diode_model_subsection}
\par Let us now abstract the rectifiers in Fig. \ref{rectenna_circuit} into the simplified representation in Fig. \ref{antenna_model} (right hand side). We consider for simplicity a rectifier composed of a single series diode followed by a low-pass filter with a load. We consider this setup as it is the simplest rectifier configuration. Nevertheless the model presented in this subsection is not limited to a single series diode but also holds for more general rectifiers with many diodes as shown in \cite{Clerckx:2017}.
\par Denote the voltage drop across the diode as $v_{\mathrm{d}}(t)=v_{\In}(t)-v_{\out}(t)$ where $v_{\In}(t)$ is the input voltage to the diode and $v_{\out}(t)$ is the output voltage across the load resistor. A tractable behavioral diode model is obtained by Taylor series expansion of the diode characteristic function
\begin{equation}
i_d(t)=i_{\mathrm{s}} \big(e^{\frac{v_{\mathrm{d}}(t)}{n v_{{\mathrm{t}}}}}-1 \big),
\end{equation}
with the reverse bias saturation current $i_{{\mathrm{s}}}$, the thermal voltage $v_{{\mathrm{t}}}$, the ideality factor $n$ assumed to be equal to $1.05$, around a quiescent operating point $v_{\mathrm{d}}(t)=a$. We have
\begin{equation}
i_{\mathrm{d}}(t)=\sum_{i=0}^{\infty }k_i' \left(v_{\mathrm{d}}(t)-a\right)^i,
\end{equation}
where $k_0'=i_{\mathrm{s}}\big(e^{\frac{a}{n v_{\mathrm{t}}}}-1\big)$ and $k_i'=i_{\mathrm{s}}\frac{e^{\frac{a}{n v_{\mathrm{t}}}}}{i!\left(n v_{\mathrm{t}}\right)^i}$, $i=1,\ldots,\infty$.
Choosing\footnote{We here assume a steady-state response and an ideal rectification. Namely the low pass filter is ideal such that $v_{\out}(t)$ is at constant DC level $v_{\out}$ (we drop the dependency on $t$). Similarly the output current is also at constant DC level $i_{\out}$.} $a=\mathbb{E}[ v_{\mathrm{d}}(t) ]=-v_{\out}$, we can write $i_d(t)=\sum_{i=0}^{\infty }k_i' v_{\In}(t)^i=\sum_{i=0}^{\infty }k_i' R_{\ant}^{i/2} y_{\rf}(t)^i$.

\par The DC current delivered to the load and the harvested DC power are then given by
\begin{equation}\label{diode_current_power}
i_{\out}=\mathbb{E}[i_{\mathrm{d}}(t)], \hspace{1cm} P_{\dc}^r=i_{\out}^2 R_{\mathrm{L}},
\end{equation}
respectively. Note that the operator $\mathbb{E}[\cdot]$ has the effect of taking the DC component of the diode current $i_{\mathrm{d}}(t)$ but also averaging over the potential randomness carried by the input signal $y_{\rf}(t)$. Indeed, in WIPT applications, $y_{\rf}(t)$ commonly carries information and is therefore changing at every symbol period due to the randomness of the input symbols it carries. This randomness due to modulation impacts the diode current $i_{\mathrm{d}}(t)$ and the amount of harvested energy, which is captured in the model by taking an expectation over the distribution of the input symbols \cite{Clerckx:2018b}.

\par In order to make the signal design tractable and get further insights, we truncate the Taylor expansion at the $n_o^{th}$ order. This leads to
\begin{equation}\label{diode_model}
i_{\out}\approx \sum_{i \hspace{0.1cm}\textnormal{even}}^{n_o} k_i' R_{\ant}^{i/2} \mathbb{E}\left[y_{\rf}(t)^i\right]
\end{equation}
where $n_o$ is an even integer with $n_o \geq 2$. The diode nonlinear model truncates the Taylor expansion at the $n_o>2$ order but retains the fundamental nonlinear behavior of the diode while the diode linear model truncates at the second order term ($n_o=2$).  Note that the rectifier characteristics $k_i'$ are a function of $a=-v_{\out}=-R_{\mathrm{L}}i_{\out}$ and therefore a function of $i_{\out}$, which makes it difficult to express $i_{\out}$ explicitly as a function of $y_{\rf}(t)$ based on \eqref{diode_model}. Fortunately, it is shown in \cite{Clerckx:2016b} that from a transmit signal optimization perspective, maximizing $i_{\out}$ in \eqref{diode_model} (subject to an RF transmit power constraint), and therefore $P_{\dc}^r$ in \eqref{diode_current_power}, is equivalent to maximizing the quantity
\begin{equation}\label{z_DC_def}
z_{\dc}=\sum_{i \hspace{0.1cm}\textnormal{even}, i\geq 2}^{n_o} k_i \mathbb{E}\left[y_{\rf}(t)^i\right]
\end{equation}
where $k_i=\frac{i_{\mathrm{s}} R_{\ant}^{i/2}}{i!\left(n v_{\mathrm{t}}\right)^i}$. Parameters $k_i$ and $z_{\dc}$ are now independent of the quiescent operating point $a$. Readers are referred to \cite{Clerckx:2016b,Clerckx:2017,Clerckx:2018b} for more details on this model.

\par The diode \textit{linear} model is obtained by truncating at order 2 such that $z_{\dc}=k_2 \mathbb{E}\left[y_{\rf}(t)^2\right]$. Under the linear model, since $k_2$ is a constant independent of the input signal, the best transmit strategy for maximizing $z_{\dc}$, subject to a transmit RF power constraint, is equivalent to the one that maximizes the average input power $P_{\rf}^r=\mathbb{E}\left[y_{\rf}(t)^2\right]$ to the rectenna \cite{Clerckx:2016b}. In other words, the diode linear model assumes that the RF-to-DC conversion efficiency $e_3$ of the rectifier is a constant independent of $y_{\rf}(t)$ \cite{Zeng:2017}. The diode linear model can therefore equivalently be written as $P_{\dc}^r= e_3 P_{\rf}^r $ with $0\leq e_3\leq 1$ a \textit{constant independent of the rectifier's input signal power and shape}.
\par
\begin{figure}
\centering
\includegraphics[scale=0.25]{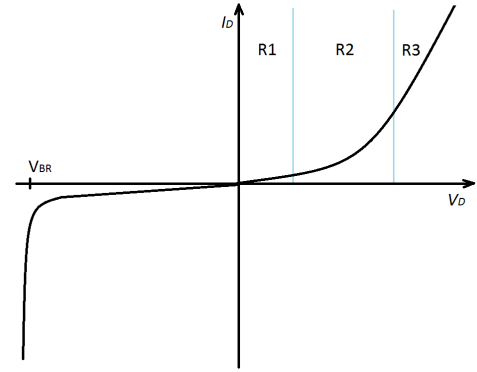}
\caption{Diode $I-V$ characteristic showing the three regions of diode operation \cite{OptBehaviour}. R1 and R2 correspond to the diode operation of the diode linear model and the diode nonlinear model, respectively. R3 corresponds to the region where the diode acts as a resistor.}
\label{diode_regions}
\end{figure}

 This is the energy harvester model first introduced in \cite{Wetenkamp:1983} and adopted in the early works on WIPT \cite{Zhou:2013}. It has since then been used extensively throughout the WIPT literature, with among others \cite{Varshney:2008,Grover:2010,Zhang:2013,Son:2014,Xu:2014,Zhou:2013,Liu:2013,Park:2013,Park:2014,Park:2015,Huang:2013,Zhou:2014,Ng:2013,Nasir:2013,Huang:2015,Ju:2014,Lee:2016,Boyer:2014,Yang:2015,Bi:2015,Tabassum:2015,KHuang:2015,Bi:2016,Krikidis:2014,Ding:2015,Chen:2015,Lu:2015}. Such a model holds whenever the higher order terms are found negligible. This occurs in the very low input power, $P_{\rf}^r$, regime or equivalently whenever the voltage drop across is the diode is small as illustrated by region R1 in Fig. \ref{diode_regions}. Such a regime is commonly denoted as the square-law regime of the diode in the RF literature \cite{OptBehaviour}. According to \cite{Minck:1997}, such a regime occurs for $P_{\rf}^r$ below -20dBm with a continuous wave (CW) input signal. When the input signal is a multisine, the higher order terms become increasingly important as the number of sinewaves increases. This has as a consequence that the square-law regime (where the diode linear model is valid) is shifted towards a lower range of average input power, namely below -30dBm \cite{Clerckx:2016b,Zeng:2017,DelPrete:2016}. Recall nevertheless that power levels below -30dBm are very low for operating state-of-the-art rectifiers since the Schottky diode is not easily turned on.

\par The diode \textit{nonlinear} model is obtained by truncating to a higher order term with $n_o\geq 4$ \cite{Clerckx:2015,Clerckx:2016b}. Choosing $n_o=4$ for simplicity, $z_{\dc}= k_2 \mathbb{E}\left[y_{\rf}(t)^2\right]+k_4 \mathbb{E}\left[y_{\rf}(t)^4\right]$ and the nonlinearity is characterized through the presence of the fourth-order term $\mathbb{E}\left[y_{\rf}(t)^4\right]$. Such a model holds whenever the higher order terms are found non-negligible. This occurs in region R2 in Fig. \ref{diode_regions}. Region R2 is often called transition region in the RF literature \cite{OptBehaviour}. The transition region ranges from $-20$ to 0 dBm average input power, when a CW input signal is considered. When using a multisine input signal, the transition region shifts to a lower range of average input powers, e.g. $[-30,-10]$dBm, as given in \cite{DelPrete:2016}. Generally speaking, the diode behavior is known in the RF literature to be highly nonlinear in the low power regime of -30dBm to 0dBm, as discussed in \cite{Clerckx:2018} and references therein.
\par For the diode nonlinear model, finding the best transmit strategy so as to maximize $z_{\dc}$, subject to an RF transmit power constraint, does not lead to the same solution as the one that maximizes $\mathbb{E}\left[y_{\rf}(t)^2\right]$. This model accounts for the dependence of the RF-to-DC conversion efficiency $e_3$ of the rectifier on the input signal (waveform shape, power, and modulation format) \cite{Clerckx:2016b,Zeng:2017}. The diode nonlinear model is a simple form of a memoryless polynomial model that has been widely adopted and validated in the RF literature \cite{Pedro:2003,Boaventura:2011,OptBehaviour}. It has since then been used in various signal design literature for WPT \cite{Huang:2017,Huang:2018,Clerckx:2017,Moghadam:2017}, SWIPT \cite{Clerckx:2016,Clerckx:2018b,Kim:2016,Bayguzina:2018,Varasteh:2017,Morsi:2017,Kang:2018,Varasteh:2017b,Varasteh:2018} and WPBC \cite{Clerckx:2017b,Zawawi:2018}.

\begin{remark} \label{remark_NL_model} As noted in \cite{Clerckx:2016b}, the Taylor series expansion around a quiescent point $a$ is a small-signal model that is valid only for the nonlinear operating region of the diode. If the input voltage amplitude becomes large, the diode will be driven into the large-signal operating region where the diode behavior is dominated by the diode series resistance and the I-V relationship is linear as illustrated by region R3 in Fig. \ref{diode_regions} \cite{OptBehaviour}.
\end{remark}

\subsection{The Saturation Nonlinear Model}\label{saturation_model_subsection}

The saturation nonlinear model characterizes another source of nonlinearity in the rectenna that originates from the saturation of the output DC power beyond a certain input RF power due to the diode breakdown\footnote{Though the term ``diode'' is not highlighted in ``saturation nonlinear model'' in contrast to the previous two models, we need to keep in mind that saturation also originates from the diode behavior.}. As illustrated in Fig. \ref{Pdc_Pin}, $e_3$ sharply decreases once the rectifier operates in the diode breakdown region\footnote{Operating diodes in the breakdown region is not the purpose of a rectifier and should be avoided as much as possible. A rectifier is designed in such a way that current flows in only one direction, not in both directions as it would occur in the breakdown region.}. The diode breakdown occurs when the diode is reversed biased with a voltage across the diode being larger than the diode breakdown voltage V\textsubscript{BR}, as illustrated in Fig. \ref{diode_regions}. At such a voltage, the breakdown is characterized by a sudden increase of the current flowing in the opposite direction (hence the negative sign of the current in Fig. \ref{diode_regions} around the breakdown voltage). This can occur typically when the input power to the rectifier is too large for the power regime it has been designed for.
\par The saturation nonlinear model is a tractable parametric model proposed in \cite{Boshkovska:2015}, and is applicable to SWIPT systems for a given pre-defined signal waveform and only based on the average received RF power $P_{\rf}^r$. Unlike the diode nonlinear model discussed in the previous subsection that is based on the physics of the diode, the nonlinear parametric saturation model is fit to measurement results obtained from practical RF-based energy harvesting circuits (excited using the pre-defined signal waveform) via curve fitting. Specifically, the total harvested power at an energy harvesting receiver, $P_{\dc}^r$, is modeled as:
 \begin{eqnarray}\label{eqn:EH_non_linear}
\hspace*{-5mm}P_{\dc}^r\hspace*{-1.0mm}&=&\hspace*{-1.0mm}
 \frac{[\Psi_{\mathrm{dc}}
 - P_{\mathrm{Sat}}\Omega]}{1-\Omega},\, \Omega=\frac{1}{1+\exp(ab)},\\
\hspace*{-2mm}\mbox{where}\,\,\Psi_{\mathrm{dc}}\hspace*{-1.0mm}&=&\hspace*{-1.0mm} \frac{P_{\mathrm{Sat}}}{1+\exp\Big(-a(P_{\rf}^r-b)\Big)}
  \end{eqnarray}
is a sigmoid (logistic) function which has the received RF power, $P_{\rf}^r$,  as input. Constant $P_{\mathrm{Sat}}$ denotes the maximal harvested power at the energy harvesting receiver when the energy harvesting circuit is driven to  \emph{saturation} due to an exceedingly large input RF power.    Constants $a$ and $b$  capture the joint effects of  resistance, capacitance, and circuit sensitivity. In particular,  $a$ reflects the nonlinear  charging rate (e.g. the steepness of the curve) with respect to the input power and $b$ determines the minimum turn-on voltage of the energy harvesting circuit.

\begin{figure}[t]
 \centering
\begin{subfigmatrix}{2}
\subfigure[Linear scale.]{\label{a}\includegraphics[width=3.5 in]{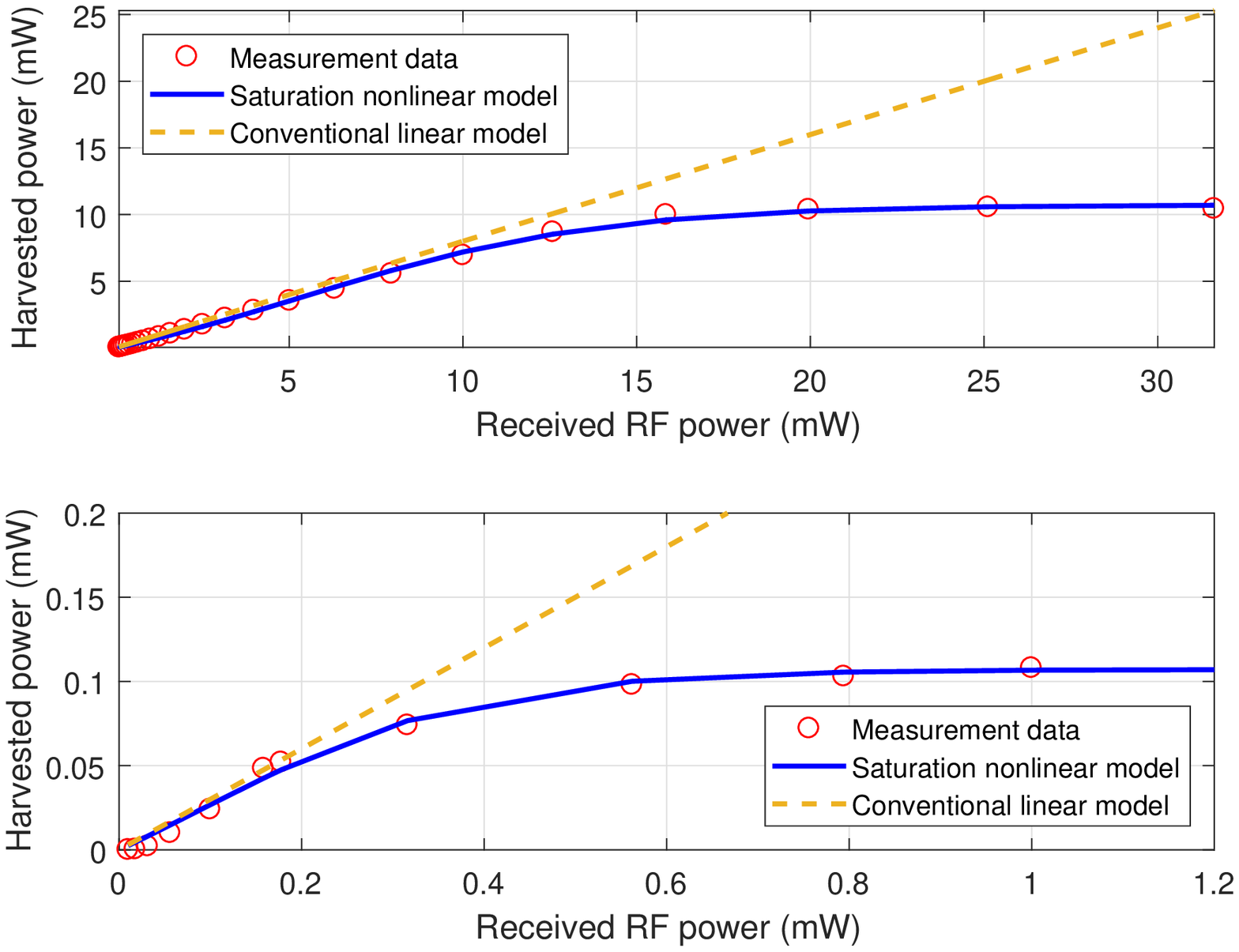}}
\subfigure[Logarithmic scale.]{\label{b}\includegraphics[width=3.5 in]{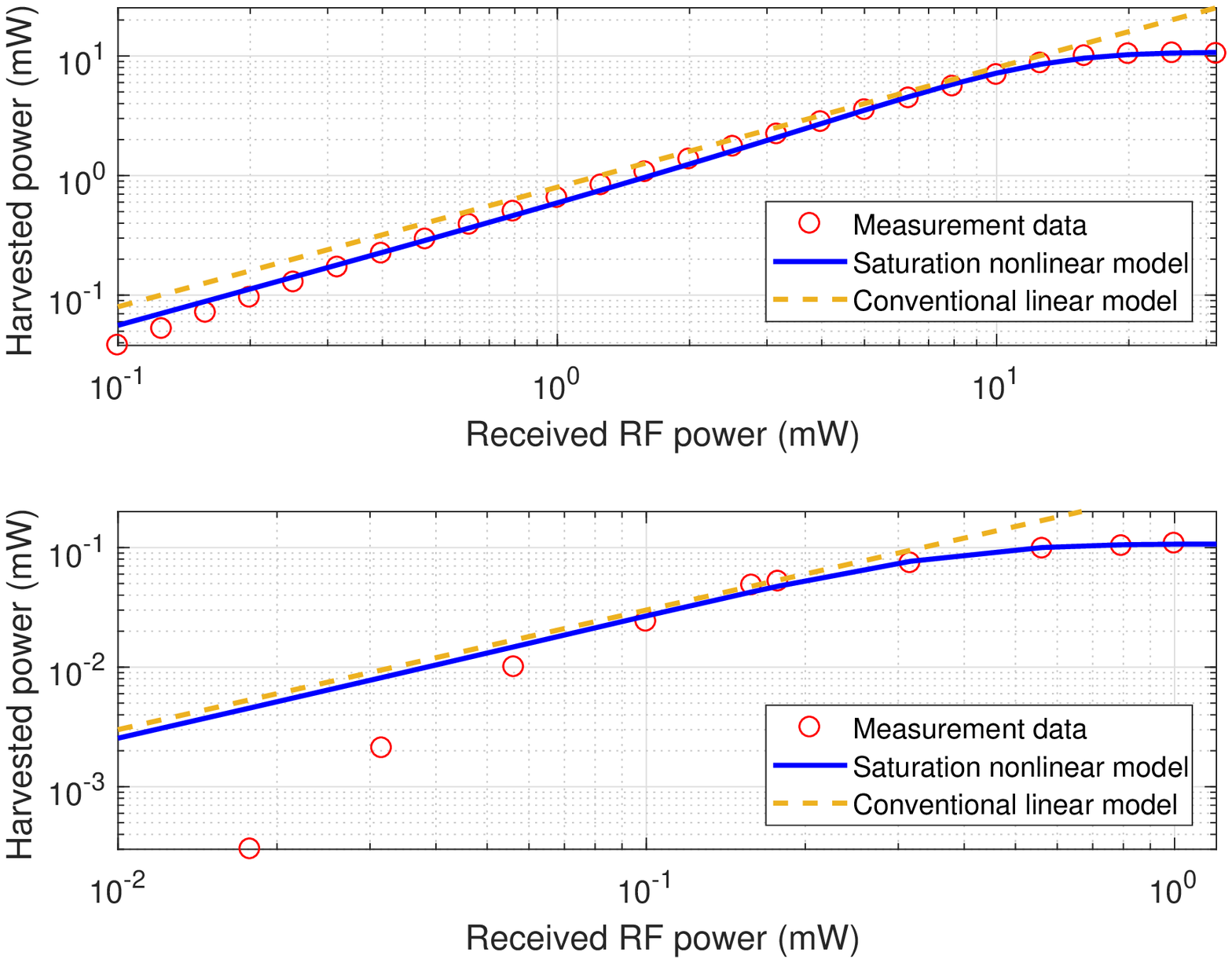}}
\end{subfigmatrix}
 \caption{ A comparison between the harvested power for the proposed model
in (\ref{eqn:EH_non_linear}) and measurement data obtained for two different practical energy harvesting circuits with a continuous wave (CW) as input signal. Fig (a) and (b) contain the same information but using two different scales, namely linear scales for (a) and logarithmic scales for (b). The measurement data of the upper and lower subfigure (for both (a) and (b)) have been taken from \cite{CN:EH_measurement_2} and \cite{JR:EH_measurement_1}, respectively, showing the different dynamic ranges in harvested energy of practical energy harvesting circuits. The parameters $a$, $b$, and $P_{\mathrm{Sat}}$ in (\ref{eqn:EH_non_linear}) are calculated with a standard curve fitting tool. The RF-to-DC conversion efficiency of the energy harvesting receiver for the linear energy harvesting model is set to $e_3=0.8$ and $e_3=0.3$ in the conventional linear model for the upper and lower subfigure, respectively.} \label{fig:comparsion_EH}
\end{figure}

\par This model isolates the resource allocation algorithm  for practical  SWIPT systems from the specific implementation details of the energy harvesting circuit and signal waveform distribution. In practice, for a given waveform of the adopted RF signal,  parameters $a$, $b$, and $P_{\mathrm{Sat}}$ of the model in (\ref{eqn:EH_non_linear}) can be obtained by applying a standard curve fitting algorithm to measurement results of a given energy harvesting hardware circuit.  In Fig. \ref{fig:comparsion_EH}, we show two examples for the curve fitting for the saturation nonlinear energy harvesting model. For the upper and lower subfigure in Fig. \ref{fig:comparsion_EH} (a) and (b), the parameters are $\{P_{\mathrm{Sat}}=10.73$ mW, $b=0.2308$,  $a=5.365\}$ and $\{P_{\mathrm{Sat}}=0.1071$ mW, $b=0.6614$,  $a=0.8963\}$,  for input powers in the mW and $10^{-4}$  W range, respectively. As can be observed, in the high power regime ($P_{\rf}^r \geq -10$ dBm$=10^{-1}$ mW), the parametric nonlinear model closely matches the experimental results provided in \cite{CN:EH_measurement_2} and \cite{JR:EH_measurement_1} for the wireless power harvested by a practical energy harvesting circuit. Fig.~\ref{fig:comparsion_EH} also illustrates the inability of the conventional (diode) linear model to capture the nonlinear characteristics of practical energy harvesting circuits in the high received RF power regime. In the low power regime, both the conventional (diode) linear model and the saturation nonlinear model experience some discrepancies. The saturation nonlinear model has been widely adopted in the literature for resource allocation algorithm design, e.g. \cite{JR:Elena_TCOM}\nocite{JR:Kim_non_linear_saturated,Kang_Kim:2017,JR:Yik_Chung_non_linear_saturated,JR:Zhou_non_linear_saturated,JR:non-linear_EH_HWI}--\cite{JR:Xiong_non_linear}.

\subsection{Comparisons of The Rectenna Models}

Table \ref{rectenna_models_comp} provides a comparison of the three models. Further comparisons between the diode linear and nonlinear models can be found in \cite{Clerckx:2016b,Clerckx:2018b,Zeng:2017}. In particular, it was observed from circuit simulations that the diode nonlinear model more accurately characterizes the rectenna behavior in the practical low power regime. For more discussions on the similarities and differences between the diode nonlienar model and the saturation nonlinear models, the readers are referred to Remark 5 in \cite{Clerckx:2018b}.

\begin{table*}
\caption{Comparisons of the rectenna models.}
\centering
\begin{tabular}{|p{2cm}||p{4.5cm}|p{4.5cm}|p{4.5cm}|}
\hline    & \textbf{Diode Linear Model}	& \textbf{Diode Nonlinear Model} &	\textbf{Saturation Nonlinear Model} \\
\hline
\hline \textbf{Operation Regime} & Characterizes the diode behavior at very low power (below -30dBm) & Characterizes the diode behavior at low power (-30dBm to 0dBm) & Characterizes the diode/rectenna behavior at high power in/around the diode breakdown region (above 0dBm) \\
\hline \textbf{$e_3$} & Constant & Function of the rectifier input signal power and shape & Function of the rectifier input signal power and shape \\
\hline \textbf{Philosophy} & Driven by simplicity & Driven by the physics of the rectenna & Curve fitting based on measured data \\
\hline \textbf{Beamforming} & Suitable for beamforming design & Suitable for beamforming design & Suitable for beamforming design\\
\hline \textbf{Modulation and Waveform} & Does not reflect dependence on input signal power and shape. Cannot be used for modulation and waveform design. & Does reflect dependence on input signal power and shape. Can be used for modulation and waveform design. & Fitted to a given pre-defined signal. Cannot be used for modulation and waveform design design. \\
\hline \textbf{Resource Allocation (RA)} & Suitable for RA optimization & Suitable for RA optimization & Suitable for RA optimization \\
\hline \textbf{Impact} & Neutral  & Diode nonlinearity is beneficial & Saturation is detrimental. Avoidable by proper (adaptive) rectifier design. \\
\hline \textbf{Rectenna} &  & Valid for rectifiers with single diode and multiple diodes & Parameters are circuit-specific \\
\hline \textbf{Applications} & For system-level performance evaluations & For PHY layer signal design and performance evaluations  & For system-level performance evaluations \\
\hline
\end{tabular}
\label{rectenna_models_comp}
\vspace{-0.3cm}
\end{table*}

\subsection{Extension and Future Work}
In the following, we review some interesting future research directions. The challenge is finding accurate but tractable models for the energy harvesters that can be used for signal and system design. Software-based models of the energy harvester exist but are insufficiently fast and not insightful to derive new signal design and optimization. Nonetheless, they are very handy when it comes to validating analytical models. On the other hand, simple models such as the linear model can be over-simplified and do not reflect the rectenna behavior accurately enough. The nonlinear models described above try to keep some level of tractability while also improving upon the accuracy compared to the linear model. Nevertheless, much remains to be done in designing rectenna model that are suited to signal and system designs. We here mention a few interesting research avenues.
\par First, we may think of developing a combined diode and saturation nonlinear model so as to tackle both sources of nonlinearity at once and cope with a wider range of input power.
\par Second, we may want to provide alternative or enhanced models for the diode and saturation nonlinearities or for the general energy harvester. Some alternative models have emerged in \cite{Varasteh:2018,Wang:2017,Xu:2017,Alevizos:2018a,Dong:2016}. In view of Fig. \ref{fig:comparsion_EH}(b), more works are also needed to better capture the harvester behavior in the low-power regime. Moreover, those models are always assuming CW input signals. It would also be beneficial to design new signals using the diode nonlinear model, validate it through circuit simulations, and then fit data using some curve fitting tool mechanism. The resulting model could then be used for system level evaluations and would capture the dependence on input signal shape and power. The sensitivity is another important characteristic of the energy harvester in the low-power regime that needs to be further investigated \cite{Alevizos:2018,Assimonis:2016}.
\par Third, we may need to consider other sources of nonlinearity in the energy harvester, such as the impedance mismatch and the rectifier output harmonics. Modeling accurately the impedance mismatch due to variation in the input signal power (accounting for fading) and shape is a challenge. Unfortunately, due to the dynamic nature of the wireless channel, the input power and signal change dynamically, implying that impedance matching cannot always be guaranteed.
\par Fourth, nonlinearities were considered at the receiver side but also exist at the transmitter side. Modeling PA nonlinearities jointly with the EH nonlinearity would result in more efficient WPT and WIPT signal designs. One way forward studied in \cite{Clerckx:2016b} consists in designing transmit signal to maximize the harvested DC power subject to an average power constraint and transmit PAPR constraints. Such a design leads to a new tradeoff since low PAPR signals are preferred at the transmitter but high PAPR signals at the input of the energy harvester.
\par Fifth, the design and modeling of energy harvester for other frequency bands, e.g. millimeter-wave band, is also of high interests. At those frequencies, the diode linear model was also shown not to accurately model the rectification behavior of the diode \cite{Ladan:2015}.

\section{Single-User WIPT}\label{single_user_WIPT_section}

In this section, we first introduce the signal model used throughout the manuscript. We then discuss various receiver architectures and formulate the R-E region maximization problem. The core part of the section is dedicated to characterizing the R-E region (and the corresponding signal design strategies) for the three energy harvester models.

\subsection{Signal and System Model}

\par We consider a single-user point-to-point MIMO SWIPT system in a general multipath environment. This setup is referred to as ``SWIPT with co-located receivers'' in Fig. \ref{SWIPT_figure}. The transmitter is equipped with $M_\mathrm{t}$ antennas that transmit information and power simultaneously to a receiver equipped with $M_\mathrm{r}$ receive antennas. We consider the general setup of a multi-subband transmission (with a single subband being a special case) employing $N$ orthogonal subbands where the $n^{\textnormal{th}}$ subband has carrier frequency $f_n$ and all subbands employ equal bandwidth $f_{\mathrm{w}}$, $n=0,...,N-1$. The carrier frequencies are evenly spaced such that $f_n=f_0+n \Delta_f$ with the inter-carrier frequency spacing $\Delta_f$ (with $f_{\mathrm{w}}\leq \Delta_f$).

\par The SWIPT signal transmitted on antenna $m$, $x_{\rf,m}(t)$, is a multi-carrier modulated waveform with frequencies $f_n$, $n=0,...,N-1$,
carrying independent information symbols on subband $n=0,...,N-1$. The transmit SWIPT signal at time $t$ on antenna $m=1,...,M_t$ is given by
\begin{equation}
x_{\rf,m}(t)=\sqrt{2}\Re\left\{\sum_{n=0}^{N-1} x_{m,n}(t) e^{j 2\pi f_n t}\right\}\label{SWIPT_WF}
\end{equation}
with the baseband equivalent signal $x_{m,n}(t)$ given by
\begin{equation}
x_{m,n}(t)=\sum_{k=-\infty}^{\infty} x_{m,n,k}\: \sinc(f_{\mathrm{w}} t-k)
\end{equation}
where $x_{m,n,k}$ denotes the complex-valued information and power carrying symbol at time index $k$, modeled as a random variable generated in an i.i.d. fashion. $x_{m,n}(t)$ has bandwidth $[-f_{\mathrm{w}}/2,f_{\mathrm{w}}/2]$.

\par The transmit SWIPT signal propagates through a multipath channel, characterized by $L$ paths. Let $\tau_l$ and $\alpha_l$ be the delay and amplitude gain of the $l^{\textnormal{th}}$ path, respectively. Further, denote by $\zeta_{i,m,n,l}$ the phase shift of the $l^{\textnormal{th}}$ path between transmit antenna $m$ and receive antenna $i$ for subband $n$. The signal received at antenna $i$ ($i=1,...,M_r$) from transmit antenna $m$ can be expressed as
\begin{align}
\!y_{\rf,i,m}(t)\!&=\!\sqrt{2}\Re\Bigg\{\sum_{l=0}^{L-1}\sum_{n=0}^{N-1} \alpha_l x_{m,n}(t-\tau_l)\Bigg. \nonumber\\
&\hspace{2.3cm}\Bigg.e^{j 2\pi f_n (t-\tau_l)+\zeta_{i,m,n,l}}\Bigg\}\!,\nonumber\\
&\approx\!\sqrt{2}\Re\left\{\sum_{n=0}^{N-1} h_{i,m,n}x_{m,n}(t) e^{j 2\pi f_n t}\right\}.\label{received_signal_ant_m}
\end{align}
We have assumed $\max_{l\neq l'}\left|\tau_l-\tau_{l'}\right|<1/f_{\mathrm{w}}$ so that, for each subband, $x_{n,m}(t)$ are narrowband signals, thus $x_{m,n}(t-\tau_l)=x_{m,n}(t)$, $\forall l$. Variable $h_{i,m,n}=\!\sum_{l=0}^{L-1}\alpha_l e^{j(-2\pi f_n\tau_l+\zeta_{i,m,n,l})}$ is the baseband channel frequency response between transmit antenna $m$ and receive antenna $i$ at frequency $f_n$.

\par The total signal and noise received at antenna $i$ is the superposition of the signals received from all $M_t$ transmit antennas, i.e.,
\begin{equation}
y_{\rf,i}(t) =  \sqrt{2}  \Re \left\{  \sum_{n=0}^{N-1} \mathbf{h}_{i,n} \mathbf x_n(t) e^{j 2\pi f_n t}  \right\}+w_{\textnormal{A},i}(t), \label{eq:yit}
\end{equation}
where $w_{\textnormal{A},i}(t)$ is the antenna noise, $\mathbf{h}_{i,n}\!\triangleq\!\big[h_{i,1,n},...,h_{i,M_t,n}\big]$ denotes the channel vector from the $M_t$ transmit antennas to receive antenna $i$ and $\mathbf{x}_{n}(t)\triangleq\big[x_{1,n}(t),...,x_{M_t,n}(t)\big]^T$ denotes the signals transmitted by the $M_t$ antennas in subband $n$. Next, the processing depends on the exact SWIPT receiver architecture. Nevertheless, a commonality exists among all considered types of receivers. Namely, from an energy perspective, $y_{\rf,i}(t)$ (or a fraction of it) is conveyed to an ER, where energy is harvested directly from the RF-domain signal. From an information perspective, $y_{\rf,i}(t)$ (or a fraction of it) is conveyed to an IR, where it is first downconverted and filtered to produce the baseband signal for subband $n$
\begin{equation}
y_{i,n}(t) =  \mathbf{h}_{i,n} \mathbf x_n(t) +w_{i,n}(t),
\end{equation}
where $w_{i,n}(t)$ is the downconverted received filtered noise, accounting for both the antenna and the RF-to-baseband processing noise. Sampling with a sampling frequency $f_{\mathrm{w}}$ to produce the sampled outputs at time instants $k$ (multiples of the sampling period), we can write the baseband system model as follows
\begin{equation}\label{BB_model_with_k}
y_{i,n,k} =  \mathbf{h}_{i,n} \mathbf x_{n,k} +w_{i,n,k}
\end{equation}
with $\mathbf x_{n,k}\triangleq\big[x_{1,n,k},...,x_{M_t,n,k}\big]^T$.
Due to the assumption of i.i.d. channel inputs and the discrete memoryless channel, we can drop the time index $k$ and simply write
\begin{equation}\label{BB_model}
y_{i,n} =  \mathbf{h}_{i,n} \mathbf x_{n} +w_{i,n}.
\end{equation}
We model $w_{i,n}$ as an i.i.d.\ and CSCG random variable with
variance $\sigma^2$, i.e., $w_{i,n}\sim\mathcal{CN}(0,\sigma^2)$, where $\sigma^2=\sigma_A^2+\sigma_{P}^2$ is the total Additive White Gaussian Noise (AWGN) power originating from the antenna ($\sigma_A^2$) and the RF-to-baseband processing ($\sigma_{P}^2$).

\par After stacking the observations from all receive antennas, we obtain
\begin{equation}\label{BB_model_vector}
\mathbf y_{n} =  \mathbf{H}_{n} \mathbf x_{n} +\mathbf w_{n},
\end{equation}
where $\mathbf y_{n}\!\triangleq\!\big[y_{1,n},...,y_{M_r,n}\big]^T$, $\mathbf w_{n}\!\triangleq\!\big[w_{1,n},...,w_{M_r,n}\big]^T$, and $\mathbf H_n \triangleq \left [\mathbf h_{1,n}^H,\cdots, \mathbf h_{M_r,n}^H\right]^H\in \mathbb{C}^{M_r\times M_t}$ denotes the MIMO channel matrix from the $M_t$ transmit antennas to the $M_r$ receive antennas at subband $n$.

\par Ignoring the noise power, the total RF power received by all $M_r$ antennas of the receiver can be expressed as
\begin{align}
P_{\rf}^r&=\sum_{i=1}^{M_r} \mathbb{E} \left[y_{\rf,i}(t)^2\right]=\sum_{i=1}^{M_r} \sum_{n=0}^{N-1} \mathbb{E} \left[|\mathbf h_{i,n} \mathbf x_n(t)|^2 \right]\notag \\
&=\sum_{n=0}^{N-1} \mathrm{Tr}\left(\mathbf H_n^H \mathbf H_n \mathbf Q_n\right),\label{eq:prfr}
\end{align}
where the positive semidefinite input covariance matrix $\mathbf Q_n$ at subband $n$ is defined as $\mathbf Q_n \triangleq \mathbb {E} \left[\mathbf x_n(t) \mathbf x_n^H(t) \right]\in \mathbb{C}^{M_t\times M_t}$. The total average transmit power is expressed as
 \begin{align}
 P_{\rf}^t=\sum_{m=1}^{M_t} \mathbb{E}[x_{\rf,m}(t)^2]= \sum_{n=0}^{N-1} \Tr(\mathbf Q_n) =\Tr(\mathbf Q),\label{eq:prft}
 \end{align}
with $\mathbf{Q}=\textnormal{diag}\left\{\mathbf{Q}_0,...,\mathbf{Q}_{N-1}\right\}$. For convenience, we also define $P_n=\Tr(\mathbf Q_n)$ as the transmit power in subband $n$. Throughout the manuscript, we will assume that the total average transmit power is subject to the constraint $P_{\rf}^t\leq P$.

\par Finally, we assume perfect Channel State Information at the Transmitter (CSIT) and perfect Channel State Information at the Receiver (CSIR).

\subsection{Receiver Architectures}

\par Various architectures for the integrated information and energy receivers in Fig. \ref{SWIPT_figure} have been proposed.

\begin{figure}
\begin{center}
\subfigure[Ideal Receiver]{\scalebox{0.5}{\includegraphics*{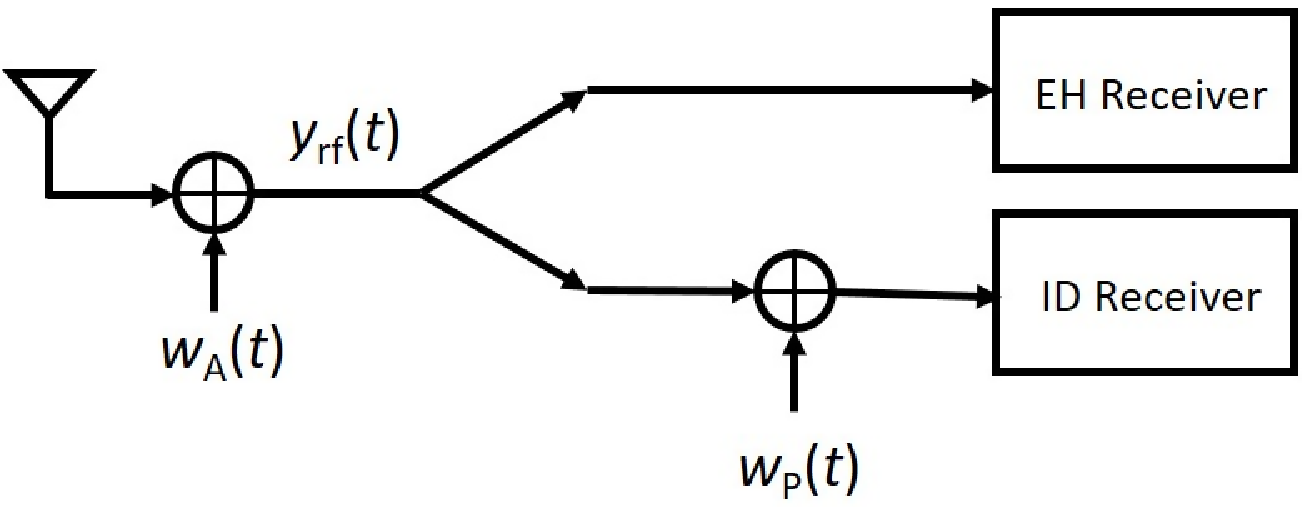}}}
\subfigure[TS Receiver]{\scalebox{0.5}{\includegraphics*{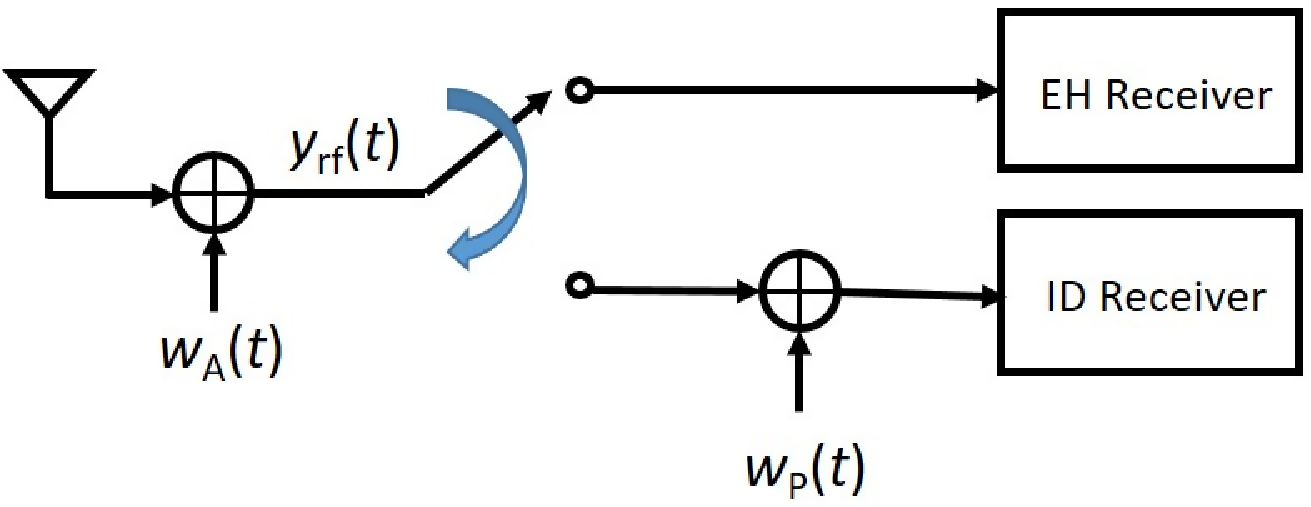}}}
\subfigure[PS Receiver]{\scalebox{0.5}{\includegraphics*{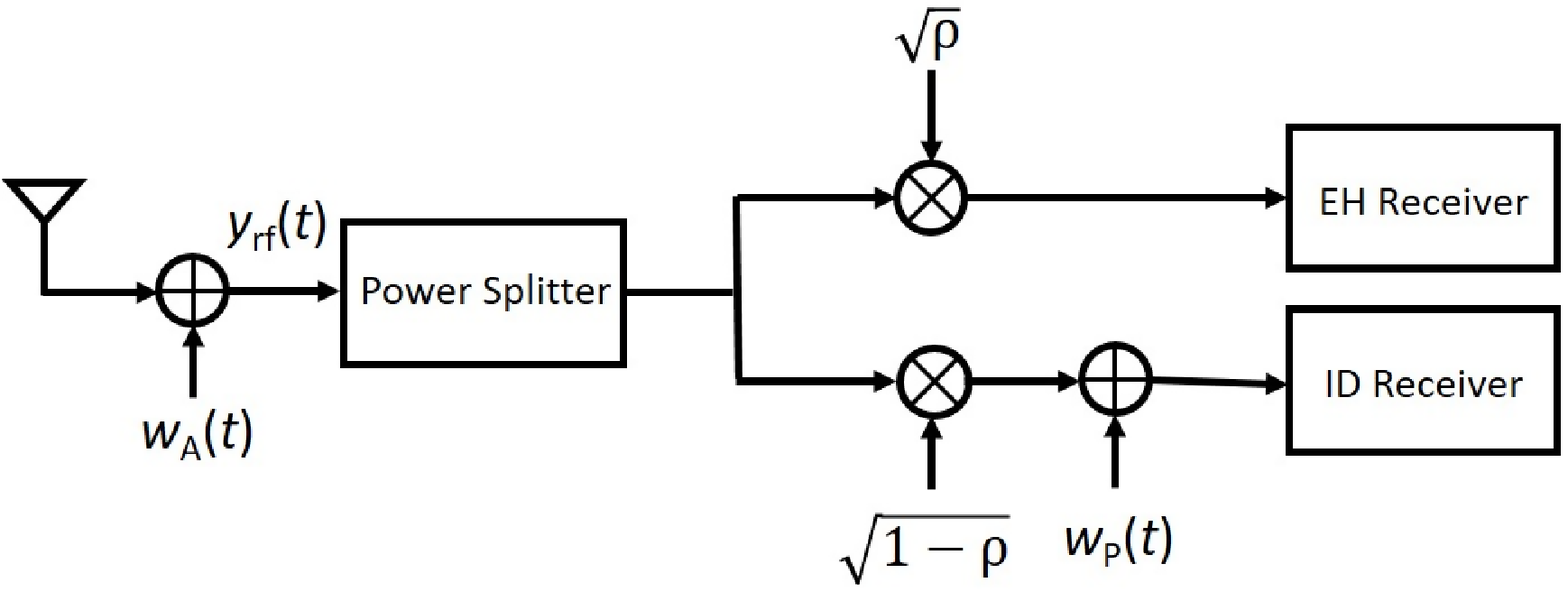}}}
\caption{Three receiver architectures for SWIPT: (a) Ideal receiver (using the same signal for both the ID and EH receivers); (b) TS receiver (switching the signal to either ID or EH receiver); and (c) PS receiver (splitting a portion of the signal to ID receiver and the rest to EH receiver).}\label{three receiver architectures}
\end{center}
\end{figure}

\par An \textit{Ideal Receiver} (Fig. \ref{three receiver architectures}(a)) is assumed to be able to decode information and harvest energy from the same signal $y_{\rf,i}(t)$ \cite{Varshney:2008,Grover:2010}; however, this cannot so far be realized by practical circuits. With such an architecture, $y_{\rf,i}(t)$ is conveyed to the energy harvester (EH) and also simultaneously RF-to-baseband downconverted and conveyed to the information decoder (ID). Different R-E tradeoffs could be realized by varying the design of the transmit signals to favor rate or energy.

\par A \textit{Time Switching (TS) Receiver} (Fig. \ref{three receiver architectures}(b)) consists of co-located ID and EH receivers, where the ID receiver is a conventional baseband information decoder; the EH receiver’s structure follows that in e.g. Fig. \ref{rectenna_circuit} \cite{Zhang:2013,Zhou:2013,Park:2013}. In this case, the transmitter divides the transmission block into two orthogonal time slots, one for transferring power and the other for transmitting data. At each time slot, the transmitter could optimize its transmit waveforms for either energy transfer or information transmission. Accordingly, the receiver switches its operation periodically between harvesting energy and decoding information in the two time slots. Then, different R-E tradeoffs could be realized by varying the length of the energy transfer time slot, jointly with the transmit signals.

\par In a \textit{Power Splitting (PS) Receiver} (Fig. \ref{three receiver architectures}(c)), the EH and ID receiver components are the same as those of a TS receiver. The transmitter optimizes the transmitted signals jointly for information and energy transfer and the PS receiver splits the received signal into two streams, where one stream with PS ratio $0 \leq \rho \leq 1$ is used for EH, and the other with power ratio $1-\rho$ is used for ID \cite{Zhang:2013,Zhou:2013,Liu:2013}. Hence, assuming perfect matching (as in Section \ref{antenna_model_subsection}), the input voltage signals $\sqrt{\rho R_{\ant}}y_{\rf}(t)$ and $\sqrt{(1-\rho)R_{\ant}}y_{\rf}(t)$ are respectively conveyed to the EH and the ID.  Different R-E tradeoffs are realized by adjusting the value of $\rho$ jointly with the transmit signals.

\subsection{Rate-Energy Region and Problem Formulation}

\par The focus of this paper is the characterization of the Rate-Energy (R-E) tradeoff and the corresponding signaling strategies for the various receiver architectures for the linear and nonlinear EH models. We define the R-E region $\mathcal{C}_{\R-\E}$ as the set of all pairs of rate $R$ and energy $E$ such that simultaneously the receiver can communicate at rate $R$ and harvested energy $E$. The R-E region in general is obtained through a collection of input distributions $p(\mathbf{x}_0,...,\mathbf{x}_{N-1})$ that satisfies the average transmit power constraint $\Tr\left(\mathbf{Q}\right)\leq P$. Mathematically, we can write
\begin{multline}\label{RE_region_def}
\mathcal{C}_{\R-\E}(P)\!\triangleq\!\bigcup_{\mycom{p(\mathbf{x}_0,...,\mathbf{x}_{N-1}):}{\Tr\left(\mathbf{Q}\right)\leq P}}\Bigg\{(R,E):R\leq \sum_{n=0}^{N-1} I\left(\mathbf{x}_n,\mathbf{y}_n\right),
\\ E\leq P_{\dc}^r\left(\mathbf{x}_0,...,\mathbf{x}_{N-1}\right) \Bigg\}
\end{multline}
where $I\left(\mathbf{x}_n,\mathbf{y}_n\right)$ refers to the mutual information between the channel input $\mathbf{x}_n$ and the channel output $\mathbf{y}_n$ on subband $n$ and $P_{\dc}^r$, function of $\mathbf{x}_0,...,\mathbf{x}_{N-1}$, refers to \eqref{diode_current_power} and \eqref{eqn:EH_non_linear} for the (linear and nonlinear) diode model and the saturation nonlinear model, respectively. For the diode models, since $P_{\dc}^r$ directly relates to the current $i_{\out}$ and therefore $z_{\dc}$ (defined in \eqref{z_DC_def}), it is more convenient to define the R-E region in terms of $z_{\dc}$, such that inequality $E\leq P_{\dc}^r$ in \eqref{RE_region_def} is replaced by $E\leq z_{\dc}$.

\par In order to characterize the R-E region, one solution is to obtain the capacity (supremization of the mutual information over all possible distributions $p(\mathbf{x}_0,...,\mathbf{x}_{N-1})$ of the input) of a complex AWGN channel subject to an average power constraint $\Tr\left(\mathbf{Q}\right)\leq P$ and a receiver delivered/harvested energy constraint $P_{\dc}^r(\mathbf{x}_0,...,\mathbf{x}_{N-1})\geq \bar{E}$, for different values of $\bar{E}\geq 0$. Namely,
\begin{align}
\sup_{p(\mathbf{x}_0,...,\mathbf{x}_{N-1})} \hspace{0.3cm}& \sum_{n=0}^{N-1}I(\mathbf{x}_n;\mathbf{y}_n) \label{MI}\\
\mathrm{subject\,\,to} \hspace{0.3cm} &\Tr\left(\mathbf{Q}\right)\leq P,\\
& P_{\dc}^r(\mathbf{x}_0,...,\mathbf{x}_{N-1})\geq \bar{E},\label{MI_2}
\end{align}
where $\bar{E}$ is interpreted as the minimum required or target harvested energy. Here again, for the diode models, it is more convenient to formulate Problem \eqref{MI}-\eqref{MI_2} in terms of $z_{\dc}$ metric such that constraint $P_{\dc}^r(\mathbf{x}_0,...,\mathbf{x}_{N-1})\geq \bar{E}$ simply writes as $z_{\dc}(\mathbf{x}_0,...,\mathbf{x}_{N-1})\geq \bar{E}$.

\par In the rest of this paper, we focus on the case when the power of the processing noise is much larger than that of the antenna noise, i.e., $\sigma_P^2\gg \sigma_A^2$, such that $\sigma^2=\sigma_A^2+\sigma_P^2\approx \sigma_P^2$. As explained in \cite{Zhang:2013}, the above setting results in the worst-case R-E region for the practical PS receiver. This can be inferred by considering the other extreme case of $\sigma_A^2\gg \sigma_P^2$ and hence $\sigma^2\approx\sigma_A^2$. In this case, it can be easily shown that the achievable rate for the ID receiver is independent of the PS ratio, and thus the optimal strategy for PS is to use an infinitesimally small split power of the received signal for ID and the remaining for EH, which achieves the same box-like R-E region (see Fig. \ref{C_threereceivers}) as the ideal receiver \cite{Zhang:2013}. As a result, we mainly consider the R-E region for the worst case of $\sigma_P^2\gg \sigma_A^2$, which serves as a performance lower bound for practical PS receivers.

\subsection{Rate-Energy Tradeoff with The Diode Linear Model}\label{L_section}

In this subsection, we study the R-E tradeoff for the diode linear model starting with the simplest case of a SISO single-subband transmission. We then extend the results to multi-subband transmission and multi-antenna transmission, before drawing some general conclusions about SWIPT signal and architecture design for the diode linear model.

\subsubsection{Single-Subband Transmission}
\begin{figure}
   \centerline{\includegraphics[width=0.9\columnwidth]{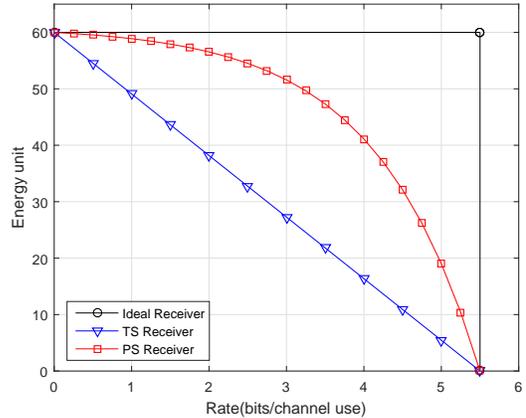}}
  \caption{Comparison of R-E tradeoffs of different SWIPT receivers with the diode linear model. The parameters are set as follows: $k_2=0.5$, the channel power is $|h|^2=12$, the transmit power is $P=10$, the noise power is $\sigma^2=3$. }
  \label{C_threereceivers}
\end{figure}

Let us first assume a SISO ($M_t=M_r=1$) single-subband ($N=1$) transmission and the ideal receiver. The system model in \eqref{BB_model} simplifies to $y =  h x +w$ and the delivered power can be expressed as $z_{\dc}(x)=k_2\mathbb{E} \left[y_{\rf}(t)^2\right]=k_2(|h|^2\mathbb{E} \left[|x|^2\right]+\sigma^2)\approx k_2|h|^2\mathbb{E} \left[|x|^2\right]$, where we assumed that the noise is negligible for energy harvesting. Problem \eqref{MI}-\eqref{MI_2} can then be written as
\begin{align}\label{SS_linear_opt}
\sup_{p(x)} \hspace{0.3cm}& I(x;y) \\
\mathrm{subject\,\,to} \hspace{0.3cm} &\mathbb{E}\left[|x|^2\right]\leq P,\\
& \mathbb{E}\left[|x|^2\right]\geq \bar{E}/(k_2|h|^2).\label{SS_linear_opt_EH_constraint}
\end{align}
Following \cite{Varshney:2008,Cover:1991}, the optimal input distribution\footnote{We here consider an average power constraint only. Under average power and amplitude constraints, the optimal capacity achieving distribution is discrete with a finite number of mass points for the amplitude
and continuous uniform over $[0,2\pi)$ for the phase \cite{Smith:1971,Shamai:1995,Varshney:2008}.} is CSCG with average transmit power $P_{\rf}^t=P$, namely $x\sim\mathcal{CN}(0,P)$, and there is no tradeoff between rate and energy, as noticed in \cite{Grover:2010}. In other words, the R-E region $\mathcal{C}_{\R-\E}^{\textnormal{L,Ideal}}$ is a rectangle characterized by \eqref{C_ideal} illustrated in Fig. \ref{C_threereceivers}.
\par For the TS and PS receivers, CSCG input is again optimal for the diode linear model. TS leads to a triangular R-E region $\mathcal{C}_{\R-\E}^{\textnormal{L,TS}}$ characterized by \eqref{C_TS} where $\tau$ is the fraction of time used for energy harvesting. PS leads to a concave-shape R-E region $\mathcal{C}_{\R-\E}^\textnormal{L,PS}$ characterized by \eqref{C_PS} where $\rho$ is the PS ratio. Hence, in the single-subband case with the diode linear model, the tradeoff between rate and energy is actually induced by the receiver architecture, not by the transmit signal.

\begin{table*}
\begin{align}
\mathcal{C}_{\R-\E}^{\textnormal{L,Ideal}}&=\Bigg\{(R,E):R\!\leq\! \log_2\!\left(1\!+\!\frac{|h|^2 P}{\sigma^2}\right), E\!\leq\! k_2 |h|^2 P \Bigg\}.\label{C_ideal}\\
\mathcal{C}_{\R-\E}^{\textnormal{L,TS}}&=\bigcup_{0\!\leq\! \tau \!\leq\! 1}\Bigg\{(R,E):R\!\leq\! \left(1-\tau\right)\log_2\left(1+\frac{|h|^2 P}{\sigma^2}\right), E\!\leq\! \tau k_2 |h|^2 P \Bigg\}.\label{C_TS}\\
\mathcal{C}_{\R-\E}^{\textnormal{L,PS}}&=\bigcup_{0\!\leq\! \rho \!\leq\! 1}\Bigg\{(R,E):R\!\leq\! \log_2\left(1+\frac{(1-\rho)|h|^2 P}{(1-\rho)\sigma_A^2+\sigma_{P}^2}\right), E\!\leq\! \rho k_2 |h|^2 P \Bigg\}.\label{C_PS}
\end{align}\hrulefill
\end{table*}

\par Comparing the three considered regions, we observe that $\mathcal{C}_{\R-\E}^{\textnormal{L,TS}}\subseteq \mathcal{C}_{\R-\E}^\textnormal{L,PS} \subseteq \mathcal{C}_{\R-\E}^{\textnormal{L,Ideal}}$. Hence, a TS receiver is outperformed by a PS receiver, and they are both outperformed by the ideal receiver. This is further illustrated in Fig. \ref{C_threereceivers}.

\subsubsection{Multi-Subband Transmission} \label{Linear_MS_section}
Let us now consider the SISO multi-subband transmission such that \eqref{BB_model} becomes $y_n =  h_n x_n +w_n$ in subband $n$. This was first investigated in \cite{Grover:2010} for the ideal receiver. Following \cite{Grover:2010}, the use of independent CSCG inputs in each subband, i.e., $x_n\sim\mathcal{CN}(0,P_n)$, is optimal and the R-E tradeoff results from the power allocation across subbands. Indeed, while the maximization of energy $\sum_{n=0}^{N-1}\left|h_n\right|^2P_n$ subject to an average sum power constraint $\sum_{n=0}^{N-1}P_n\leq P$ favors allocating all power to a single subband, namely the one corresponding to the strongest channel $\max_{n\in\{0,\ldots,N-1\}}|h_n|$, the maximization of rate subject to an average sum power constraint in general allocates power to multiple subchannels following the standard water-filling (WF) solution \cite{Cover:1991}. Hence, there exists a non-trivial tradeoff between rate and energy in the multi-subband case and the best power allocation can be formulated as the solution of the optimization problem
\begin{align}
\max_{\left\{P_0,...,P_{N-1}\right\}} \hspace{0.3cm}& \sum_{n=0}^{N-1} \log_2\left(1+\frac{\left|h_n\right|^2P_n}{\sigma^2} \right) \label{multicarrier_opt_problem}\\
\mathrm{subject\,\,to} \hspace{0.3cm} &\sum_{n=0}^{N-1}P_n\leq P,\label{sum-power constraint}\\
& \sum_{n=0}^{N-1}\left|h_n\right|^2P_n\geq \bar{E}/k_2,\label{harvested-energy constraint}
\end{align}
which yields a modified WF solution \cite{Grover:2010}.
Specifically, let $\lambda^{\star}$ and $\beta^{\star}$ denote the optimal dual variables corresponding to the transmit sum-power constraint (\ref{sum-power constraint}) and the total harvested power constraint (\ref{harvested-energy constraint}). Then, the optimal transmit power allocation is given by \cite{Grover:2010}
\begin{align}\label{modified water filling}
P_n^{\star}=\max\left(\frac{1}{\lambda^{\star}-\beta^{\star}|h_n|^2}-\frac{\sigma^2}{|h_n|^2},0\right),
\end{align}
$\forall n\!\in\!\left\{0,...,N\!-\!1\right\}$. It can be observed that if the energy harvesting constraint \eqref{harvested-energy constraint} is not active, i.e., $\beta^{\star}=0$, (\ref{modified water filling}) reduces to the conventional WF power allocation with a constant water-level for all subbands. However, when the energy harvesting constraint is tight, i.e., $\beta^{\star}>0$, the water-level is higher on subbands with stronger channel power. This indicates that due to the energy harvesting constraint, the power allocation among subbands is more greedy (i.e., more power is assigned to stronger subbands) than the conventional WF power allocation.

\par The TS architecture relies on time-sharing between conventional WF (for rate maximization) and transmission on the strongest subband (for energy maximization). In the PS architecture, the PS ratio (same for all the subbands) and the power allocations across subbands can be jointly optimized \cite{Zhou:2014}. Similarly to the single-subband case, $\mathcal{C}_{\R-\E}^{\textnormal{L,TS}}\subseteq \mathcal{C}_{\R-\E}^\textnormal{L,PS} \subseteq \mathcal{C}_{\R-\E}^{\textnormal{L,Ideal}}$ also holds for the multi-subband case. In fact, this result can be obtained from \cite{Zhang:2013}, which considers the general MIMO system model $\mathbf y =  \mathbf{H} \mathbf x +\mathbf w$ (see next subsection for more details). As shown in \cite{Zhang:2013}, for arbitrary MIMO channel matrix $\mathbf H$, under the so-called uniform power splitting (UPS) scheme, in which the PS ratios in each dimension of $\mathbf y$ are identical, the corresponding R-E region is always no smaller than that achieved by applying TS in each dimension of $\mathbf y$. As a result, in a multi-subband SISO system, $\mathcal{C}_{\R-\E}^{\textnormal{L,TS}}\subseteq \mathcal{C}_{\R-\E}^\textnormal{L,PS}$ follows directly by restricting $\mathbf H$ in \cite{Zhang:2013} to an $N$-by-$N$ diagonal channel.

\par Hence, in the multi-subband case for the diode linear model, a tradeoff between rate and energy is induced by the power allocation strategy at the transmitter additionally to the tradeoff already induced by the receiver architecture (as in the single-subband case).

\subsubsection{Multi-Antenna Transmission}\label{Linear_MA_section}
Let us now consider a MIMO transmission and assume a single subband for simplicity such that \eqref{BB_model_vector} becomes $\mathbf y =  \mathbf{H} \mathbf x +\mathbf w$. Similarly to the SISO case, following \cite{Telatar:1999}, the maximization of the mutual information subject to average transmit power and received power constraints is achieved by CSCG inputs. Problem \eqref{MI}-\eqref{MI_2} becomes
\begin{align}
\max_{\mathbf{Q}} \hspace{0.3cm}& \log_2\det\left(\mathbf{I}+\mathbf{H}\mathbf{Q}\mathbf{H}^H\right) \label{MI_MIMO}\\
\mathrm{subject\,\,to} \hspace{0.3cm} &\Tr\left(\mathbf{Q}\right)\leq P,\\
& \Tr\left(\mathbf{H}\mathbf{Q}\mathbf{H}^H\right)\geq \bar{E}/k_2.\label{MI_2_MIMO}
\end{align}
In the above problem formulation, we assume that each receive antenna is equipped with an energy harvester and the constraint $k_2\Tr\left(\mathbf{H}\mathbf{Q}\mathbf{H}^H\right)\geq \bar{E}$ reflects that the total harvested energy across all rectennas should be larger than $\bar{E}$.
The choice of the input covariance $\mathbf{Q}$ leads to a non-trivial R-E tradeoff \cite{Zhang:2013}. Let us write the eigenvalue decomposition $\mathbf{H}^H \mathbf{H}=\mathbf{V}_{\mathbf{H}}\mathbf{\Gamma}_{\mathbf{H}}\mathbf{V}_{\mathbf{H}}^H$. The harvested energy is maximized by choosing the covariance matrix as $\mathbf{Q}=P\mathbf{v}_1\mathbf{v}_1^H$ where $\mathbf{v}_1=\mathbf{v}_{\textnormal{max}}(\mathbf{H}^H \mathbf{H})$ denotes the eigenvector corresponding to the dominant eigenvalue of $\mathbf H^H \mathbf H$. Rate maximization on the other hand is obtained through multiple eigenmode transmission (spatial multiplexing) along the eigenvectors of $\mathbf{H}^H \mathbf{H}$ and with a power allocation across eigenmodes based on the conventional MIMO WF solution \cite{Telatar:1999}, i.e., leading to a covariance matrix of the form $\mathbf{Q}=\mathbf{V}_{\mathbf{H}}\mathbf{\Lambda}\mathbf{V}_{\mathbf{H}}^H$ with the diagonal matrix $\mathbf{\Lambda}$ obtained from the standard WF power allocation solution. The optimal solution of the R-E region maximization problem \eqref{MI_MIMO}-\eqref{MI_2_MIMO} can also be expressed in form of a multiple eigenmode transmission with $\mathbf{Q}^{\star}=\mathbf{V}_{\mathbf{H}}\mathbf{\Sigma}\mathbf{V}_{\mathbf{H}}^H$, where the diagonal matrix $\mathbf{\Sigma}$ is obtained from a modified WF power allocation \cite{Zhang:2013}. As explained in Section \ref{Linear_MS_section}, the above optimal precoder design with the modified WF power allocation is more general than the optimal power allocation for the multi-subband SISO system considered in \cite{Grover:2010}, since the channel model in \cite{Grover:2010} is a special case of that considered in \cite{Zhang:2013} with $\mathbf H$ being diagonal.

\par The TS architecture relies on time-sharing between the conventional eigenmode transmission (for rate maximization) and aligning one energy beam towards the eigenvector corresponding to the strongest eigenvalue of $\mathbf H$ (for energy maximization) \cite{Zhang:2013}. In contrast, with the PS architecture, the transmit precoder and PS ratios of receive antennas can be jointly optimized to obtain various points on the boundary of the achievable R-E region. Moreover, as mentioned in Section \ref{Linear_MS_section}, a low-complexity UPS scheme is considered in \cite{Zhang:2013} under which the PS ratios are identical for all receive antennas. Let $\mathcal{C}_{\R-\E}^\textnormal{L,UPS}$ denote the corresponding R-E region. Then, it follows from \cite{Zhang:2013} that $\mathcal{C}_{\R-\E}^{\textnormal{L,TS}}\subseteq \mathcal{C}_{\R-\E}^\textnormal{L,UPS}\subseteq \mathcal{C}_{\R-\E}^\textnormal{L,PS}\subseteq \mathcal{C}_{\R-\E}^\textnormal{L,Ideal}$.

\par Note that in the MISO setup ($M_r=1$), $y =  \mathbf{h} \mathbf x +w$, $\mathbf{Q}^{\star}=P\bar{\mathbf{h}}^H\bar{\mathbf{h}}$ with $\bar{\mathbf{h}}=\mathbf{h}/\left\|\mathbf{h}\right\|$ and the optimal covariance matrix for energy and rate maximization coincide. The transmitter simply performs conventional Maximum Ratio Transmission (MRT), $\mathbf x=\bar{\mathbf{h}}^H x$ with $x\sim\mathcal{CN}(0,P)$, which maximizes both energy and rate. Hence, there is no R-E tradeoff and the R-E region $\mathcal{C}_{\R-\E}^{\textnormal{Ideal}}$ is a rectangle characterized by \eqref{C_ideal} with $|h|^2$ replaced by $\left\|\mathbf{h}\right\|^2$, and therefore enlarged compared to the SISO case thanks to the beamforming gain. Similarly, for the TS and PS receivers, the R-E regions are given by \eqref{C_TS} and \eqref{C_PS}, respectively, with $|h|^2$ replaced by $\left\|\mathbf{h}\right\|^2$.

\begin{remark}\label{remark_linear} Note that while CSCG is optimal for the ideal, TS, and PS receivers in single-subband, multi-subband and multi-antenna transmissions for maximizing the R-E region under the diode linear model, from an energy maximization-only perspective, any input distribution with an average power $P$ would be optimal. In particular a continuous wave (CW) would do as well as a CSCG input while modulated and unmodulated waveforms are equally suitable from an energy maximization perspective under the diode linear model. Hence, in TS, the R-E region can also be achieved by time sharing with CSCG inputs during the information transmission phase and with CW during the power transmission phase.
\end{remark}

\begin{observation}\label{observation_linear} The use of the diode linear model leads to three important observations. \textit{First}, the strategy that maximizes $P_{\rf}^r$ maximizes $P_{\dc}^r$. \textit{Second}, CSCG inputs are sufficient and optimal to achieve the R-E region boundaries. \textit{Third}, $\mathcal{C}_{\R-\E}^{\textnormal{L,TS}}\subseteq \mathcal{C}_{\R-\E}^\textnormal{L,PS}\subseteq \mathcal{C}_{\R-\E}^\textnormal{L,Ideal}$.
\end{observation}

\subsection{Rate-Energy Tradeoff with The Diode Nonlinear Model}\label{NL_section}

The first systematic signal designs for WPT accounting for the diode nonlinearity appeared in \cite{Clerckx:2015,Clerckx:2016b}. Uniquely, this nonlinearity was shown to be beneficial for system performance and be exploitable (along with a beamforming gain and a channel frequency diversity gain) through suitable signal designs. It was observed that signals designed accounting for the diode nonlinearity are more efficient than those designed based on the diode linear model. Interestingly, while the diode linear model favours narrowband transmission with all the power allocated to a single subband (as  in Section \ref{Linear_MS_section}), the diode nonlinear model favours a power allocation over multiple subbands and those with stronger frequency-domain channel gains are allocated more power. The optimum power allocation strategy results from a compromise between exploiting the diode nonlinearity and the channel frequency diversity.

\par The works \cite{Clerckx:2015,Clerckx:2016b} assumed deterministic multisine waveforms. Designing SWIPT requires the transmit signals to carry information and therefore to be subject to some randomness. This raises an interesting question: How do modulated signals perform in comparison to deterministic signals for energy transfer? Recall from Remark \ref{remark_linear} that modulated and unmodulated inputs are equally suitable for energy maximization according to the diode linear model. Interestingly, it was shown in \cite{Clerckx:2018b} that modulation using CSCG inputs leads to a higher harvested energy at the output of the rectifier compared to an unmodulated input, despite exhibiting the same average power at the input to the rectenna. This gain comes from the large fourth order moment offered by CSCG inputs, which is exploited by the rectifier nonlinearity and modeled by the fourth order term in $z_{\dc}$. Indeed with CSCG inputs $x\sim\mathcal{CN}(0,P)$, $\mathbb{E}\left[|x|^4\right]=2 P^2$, which is twice as large as what is achieved with unmodulated CW inputs with the same average power \cite{Clerckx:2018b}.

\par This highlights that the signal theory and design for SWIPT, such as modulation, waveform, and input distribution, are actually influenced by the diode nonlinearity. This motivates the design of SWIPT signals that intelligently exploit the diode nonlinearity.

\subsubsection{Single-Subband Transmission}
Assuming a SISO transmission and the ideal receiver, Problem \eqref{MI}-\eqref{MI_2} becomes
\begin{align}
\sup_{p(x)} \hspace{0.3cm}& I(x;y) \label{MI_NL}\\
\mathrm{subject\,\,to} \hspace{0.3cm} &\mathbb{E}\left[|x|^2\right]\leq P,\label{MI_1_NL}\\
& k_2\mathbb{E}\left[y_{\rf}(t)^2\right]+k_4\mathbb{E}\left[y_{\rf}(t)^4\right]\geq \bar{E}.\label{MI_2_NL}
\end{align}
As shown in \cite{Varasteh:2017}, a baseband equivalent for $z_{\dc}$ (left hand side of \eqref{MI_2_NL}) is not only a function of $\mathbb{E}\left[|x|^2\right]$ as with the diode linear model but also a function of $\mathbb{E}\left[|x|^4\right]$, $\mathbb{E}\left[\Re\left\{x\right\}^\alpha\right]$, $\mathbb{E}\left[\Im\left\{x\right\}^\alpha\right]$ with $\alpha=1,2,3,4$.
\par
\begin{figure}
   \centerline{\includegraphics[width=\columnwidth]{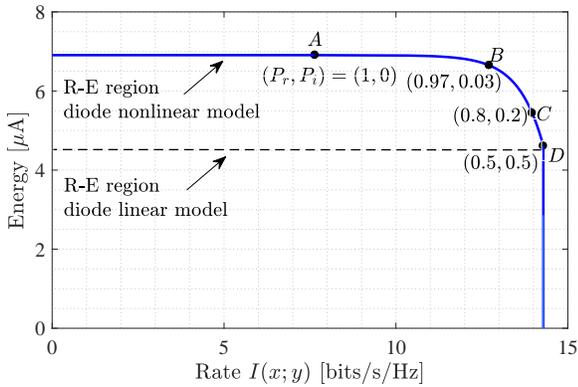}}
  \caption{R-E region with asymmetric Gaussian inputs in single-subband transmission ($P=1, \sigma^2=10^{-4}, k_2=0.17, k_4=19.145$) \cite{Varasteh:2017}. By evolving from point D to point A, the input distribution becomes more asymmetric and the harvested energy increases. The dashed line corresponds to the R-E region for the diode linear model. Note that the energy unit is $\mu$A because the energy metric used is $z_{\dc}$, which is a contribution to the output current.}
  \label{RE_region_gaussian}
\end{figure}

 Interestingly, the presence of the higher moments of the input distribution has a significant impact on the choice of the input distribution $p(x)$. In \cite{Varasteh:2017}, assuming general non-zero mean Gaussian inputs $\Re\left\{x\right\}\!\sim\!\mathcal{N}(\mu_r,P_r)$ and $\Im\left\{x\right\}\!\sim\!\mathcal{N}(\mu_i,P_i)$ with $P_r+P_i \leq P$, it is found that the supremum in Problem \eqref{MI_NL}-\eqref{MI_2_NL} is achieved with $P_r+P_i = P$ and by zero mean asymmetric Gaussian inputs. CSCG input obtained by equally distributing power between the real and the imaginary dimensions, i.e., $\Re\left\{x\right\}\!\sim\!\mathcal{N}(0,P/2)$ and $\Im\left\{x\right\}\!\sim\!\mathcal{N}(0,P/2)$ is optimal for rate maximization. However, as the harvested power constraint $\bar{E}$ increases, the input distribution becomes asymmetric with the power allocated to the real part $P_r$ increasing and the one to the imaginary part $P_i=P-P_r$ decreasing (or inversely) up to a point where the rate is minimized and the energy is maximized by allocating the transmit power to only one dimension, e.g. $\Re\left\{x\right\}\!\sim\!\mathcal{N}(0,P)$. This is because allocating power to one dimension leads to a higher fourth moment. Indeed, $\mathbb{E}\left[x^4\right]=3 P^2$ for $x\sim\mathcal{N}(0,P)$ in contrast to $\mathbb{E}\left[|x|^4\right]=2 P^2$ with $x\sim\mathcal{CN}(0,P)$. The R-E region obtained with asymmetric Gaussian inputs is illustrated in Fig. \ref{RE_region_gaussian}.
\par Hence, in contrast with the diode linear model, a R-E tradeoff exists in SISO single-subband transmission with an ideal receiver for the diode nonlinear model. The tradeoff is induced by the presence of the fourth moment of the received signal $y_{\rf}(t)$ in $z_{\dc}$. Moreover, the R-E region achieved by the diode nonlinear model-motivated input distribution leads to an enlarged R-E region compared to that achieved by the diode linear model-motivated input distribution. In other words, the diode nonlinearity is beneficial to SWIPT system performance if properly exploited.

\par
\begin{figure}
   \centerline{\includegraphics[width=\columnwidth]{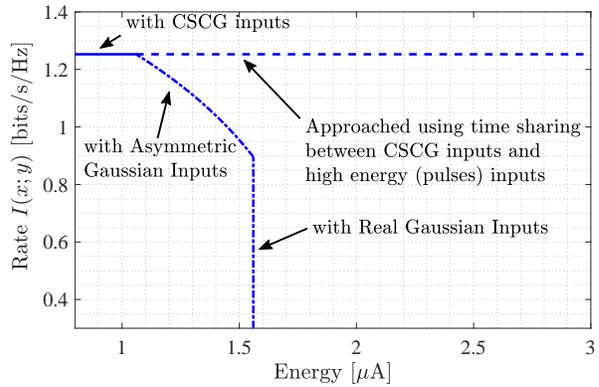}}
  \caption{R-E region with optimal inputs in single-subband transmission ($P=5, \sigma^2=2$) \cite{Varasteh:2017b}. Note that the energy unit is $\mu$A because the energy metric used is $z_{\dc}$, which is a contribution to the output current.}
  \label{RE_region_optimal}
\end{figure}

 Relaxing the constraints on Gaussian inputs, it is remarkably shown in \cite{Varasteh:2017b} that the capacity of an AWGN channel under transmit average power and receiver delivered power constraints as characterized by Problem \eqref{MI_NL}-\eqref{MI_2_NL} is actually the same as the capacity of an AWGN channel under an average power constraint, i.e., characterized by Problem \eqref{MI_NL}-\eqref{MI_1_NL} without constraint \eqref{MI_2_NL}. In other words, the capacity of an AWGN channel is independent of the value of the delivered power constraint $\bar{E}$ and the R-E region $\mathcal{C}_{\R-\E}^{\textnormal{NL,Ideal}}$ is an unbounded rectangle characterized by
\begin{equation}
\mathcal{C}_{\R-\E}^{\textnormal{NL,Ideal}}=\Bigg\{(R,E):R\!\leq\! \log_2\left(1+\frac{|h|^2 P}{\sigma^2}\right), E\!\leq\! \infty \Bigg\}.\label{C_NL_ideal}\\
\end{equation}
However, depending on the transmit average power and the receiver delivered power constraints, the capacity can be either achieved or arbitrarily approached, as illustrated in Fig. \ref{RE_region_optimal}.

Let $E_\mathrm{G}$ denote the harvested energy with the input $x\sim\mathcal{CN}(0,P)$. For $\bar{E}\leq E_\mathrm{G}$, the capacity is achieved via the unique input $x\sim\mathcal{CN}(0,P)$. For $\bar{E} > E_\mathrm{G}$, the capacity is approached by using time sharing between distributions with high amount of information, e.g.\ CSCG inputs, and distributions with high amount of power, reminiscent of flash signaling, exhibiting a low probability of high amplitude signals. Writing the complex input as $x=r e^{j\theta}$ with its phase $\theta$ uniformly distributed over $\left[0,2\pi\right)$, an example of such a flash signaling distribution is given by the following probability mass function
\begin{align}\label{flash_distr}
p_r(r)=\begin{cases}
1-\frac{1}{l^2}, \ & r=0,\\
\frac{1}{l^2}, \ & r=l\sqrt{P},
\end{cases}
\end{align}
with $l\geq 1$. We can easily verify that $\mathbb{E} \left[|x|^2\right]=\mathbb{E}\left[r^2\right]=P$, hence satisfying the average power constraint. By increasing $l$, $r=l\sqrt{P}$ increases and $p_r(l\sqrt{P})$ decreases, therefore exhibiting a low probability of high amplitude signals. Such a distribution is characterized by the fact that there is always an $L$ such that for $l\geq L$, the delivered power constraint is satisfied.

\par The benefits of departing from Gaussian inputs originate from the diode nonlinearity that favors the use of distributions that boosts the higher order moment statistics of the channel input $x$. Indeed, $\mathbb{E}\left[|x|^4\right]=l^2 P^2$ for the input distribution in \eqref{flash_distr}. Choosing $l>\sqrt{3}$ makes the fourth order moment higher than the $2P^2$ and $3P^2$ obtained respectively with real Gaussian and CSCG inputs.

\par The above discussion has deep consequences for the choice of the receiver architecture. Though the results were obtained assuming an ideal receiver, the capacity was shown to be achieved/approached with time sharing. Hence this implies that under the diode nonlinear model, a TS receiver is actually sufficient to approach the capacity. Actually, the optimal R-E regions achieved by the TS, PS, and ideal receivers are the same, i.e. $\mathcal{C}_{\R-\E}^{\textnormal{NL,TS}} = \mathcal{C}_{\R-\E}^\textnormal{NL,PS} = \mathcal{C}_{\R-\E}^{\textnormal{NL,Ideal}}$. This also results in the fact that $\mathcal{C}_{\R-\E}^{\textnormal{L,TS}}\subseteq \mathcal{C}_{\R-\E}^\textnormal{L,PS} \subseteq \mathcal{C}_{\R-\E}^{\textnormal{L,Ideal}}\subseteq\mathcal{C}_{\R-\E}^{\textnormal{NL,TS}} = \mathcal{C}_{\R-\E}^\textnormal{NL,PS} = \mathcal{C}_{\R-\E}^{\textnormal{NL,Ideal}}$, which again highlights that the diode nonlinearity is beneficial to SWIPT system performance if properly exploited. Nevertheless, in practice, the optimal distribution (resulting from the use of flash signaling) leads to high amplitude signals which may not be practical\footnote{This also calls for introducing an additional amplitude constraint in Problem \eqref{MI_NL}-\eqref{MI_2_NL}. It was shown in \cite{Varasteh:2017b} that under average power, amplitude, and nonlinear delivered power constraints, the optimal capacity achieving distributions are discrete with a finite number of mass points for the amplitude and continuous uniform for the phase.}. Given two fixed distributions, one having high information content and the other having high power content, an ideal receiver would lead to a larger R-E region than a PS receiver, which itself has a larger R-E region than a TS receiver.

\par The above discussion is illustrated in Fig. \ref{RE_region_optimal}, where the solid line illustrates the capacity achievable by $x\sim\mathcal{CN}(0,P)$ and the dashed line illustrates the capacity that can be approached arbitrarily using time sharing between distributions with high amount of information and distributions with high amount of power. Comparison is also made with the R-E region achieved with asymmetric Gaussian inputs.

\begin{remark} It is important to recall that the above observations hold as long as the conditions expressed in Remark \ref{remark_NL_model} are valid. If the signal amplitude becomes very large, the diode is forced into its resistive zone as described in \cite{OptBehaviour} (where the I-V characteristic is linear), making the Taylor series expansion model no longer applicable\footnote{The same is true for the multisine waveform in WPT with an increasing number of sinewaves \cite{Clerckx:2016b}.}. This implies that the unbounded rectangular R-E region in Fig. \ref{RE_region_optimal} cannot be achieved in practice. Nevertheless, the insights obtained from Fig. \ref{RE_region_optimal} still hold and time sharing between CSCG inputs and the distribution in \eqref{flash_distr} should enlarge the R-E region as $l$ increases up to a certain limit.
\end{remark}
	
\subsubsection{Multi-Subband Transmission}\label{MS_section_NL_SU}

\par For multi-subband transmission, the capacity achieving input distribution remains an open problem. Nevertheless, it has been shown in \cite{Clerckx:2016,Clerckx:2018b} that the use of non-zero mean Gaussian inputs leads to an enlarged R-E region compared to CSCG inputs. The superiority of non-zero mean inputs over zero mean inputs can be explained by the fact that modulated and unmodulated multi-carrier waveforms are not equally suitable for wireless power delivery. Indeed a multi-carrier unmodulated waveform, e.g. multisine, is more efficient in exploiting the diode nonlinearity and therefore boosting $z_{\dc}$ compared to a multi-carrier modulated waveform with CSCG inputs. It was indeed shown in \cite{Clerckx:2016b,Clerckx:2018b} from analysis and circuit simulations that $z_{\dc}$ scales linearly with $N$ for an unmodulated multisine waveform because all carriers are in phase, which enables the excitation of the rectifier (and the turning on of the diode) in a periodic manner by sending high energy pulses every $1/\Delta_f$. On the other hand, a modulated waveform leads to a $z_{\dc}$ that scales at most logarithmically with $N$ due to the independent CSCG randomness (and therefore random fluctuations of the amplitudes and phases) of the information-carrying symbols across subbands.

\par Non-zero mean Gaussian inputs lead to a SWIPT architecture relying on the superposition of two waveforms at the transmitter: a power waveform comprising a deterministic (unmodulated) multisine and an information waveform comprising multi-carrier modulated (with CSCG inputs) waveforms. The complex-valued information-power symbol on subband $n$ can then be explicitly written as $x_n=x_{\mathrm{P},n}+x_{\mathrm{I},n}$ with $x_{\mathrm{P},n}$ representing the deterministic power symbol of the multisine waveform on subband $n$ and $x_{\mathrm{I},n}\sim\mathcal{CN}(0,P_{\mathrm{I},n})$ representing the CSCG distributed information symbol of the modulated waveform on subband $n$. Hence, $x_n$ is non-zero mean and $|x_n|$ is Ricean distributed with a K-factor on subband $n$ denoted and given by $K_{n}=P_{\mathrm{P},n}/P_{\mathrm{I},n}$ with $P_{\mathrm{P},n}=|x_{\mathrm{P},n}|^2$.

\par Since $x_{\mathrm{P},n}$ is deterministic, the differential entropies of $x_n$ and $x_{\mathrm{I},n}$ are identical (because translation does not change the differential entropy) and the achievable rate is equal to $I\!=\!\sum_{n=0}^{N-1} \log_2\left(1+|h_n|^2 P_{\mathrm{I},n}/\sigma^2 \right)$ independent of $x_{\mathrm{P},n}$. This rate is achievable with and without waveform cancellation. In the former case, after down-conversion from RF-to-baseband (BB) and Analog-to-Digital Conversion (ADC), the contribution of the power waveform is subtracted from the received signal. In the latter case, a power waveform cancellation operation is not needed and the baseband receiver decodes the translated version of the symbols.

\par In \cite{Clerckx:2018b}, the characterization of an achievable R-E region was conducted by performing an energy maximization subject to a rate constraint
\begin{align}
\max_{\left\{x_{\mathrm{P},n},P_{\mathrm{I},n}\right\}_{\forall n}} \hspace{0.3cm}&k_2\mathbb{E}\left[y_{\rf}(t)^2\right]+k_4\mathbb{E}\left[y_{\rf}(t)^4\right]\\
\mathrm{subject\,\,to} \hspace{0.3cm} &\sum_{n=0}^{N-1}P_{\mathrm{P},n}+P_{\mathrm{I},n}\leq P,\\
& \sum_{n=0}^{N-1} \log_2\left(1+\frac{|h_n|^2 P_{\mathrm{I},n}}{\sigma^2} \right)\geq \bar{R},
\end{align}
where $\bar{R}$ denotes the minimum required rate. A similar problem can be formulated for the PS receiver where the optimization is conducted jointly over the PS ratio $\rho$ and the input variables $x_{\mathrm{P},n}$ and $P_{\mathrm{I},n}$.

\par The phases of $x_{\mathrm{P},n}$ can easily be found in closed form, while the magnitude/power terms $P_{\mathrm{P},n}$ and $P_{\mathrm{I},n}$ for the power and information symbols are found as a solution of a Reversed Geometric Program \cite{Clerckx:2018b}. The minimum energy and maximum rate is obtained by allocating no power (though energy is still harvested from the information symbols) to the deterministic power symbols, i.e.\ $P_{\mathrm{P},n}=0$ $\forall n$, and allocating power across subbands to the information symbols according to the standard WF solution. Hence, $x_n\sim\mathcal{CN}(0,P_{n})$ is CSCG and $K_{n}=0$, $\forall n$, at this extreme point. The other extreme point of the region corresponds to the maximum energy and minimum rate that is obtained by allocating no power to the information symbols, i.e., $P_{\mathrm{I},n}=0$, $\forall n$, and all power to the power symbols according to the optimal multisine waveform power allocation strategy of \cite{Clerckx:2016b}. Hence, $x_n=x_{\mathrm{P},n}$ is deterministic and $K_{n}=\infty$, $\forall n$, at this other extreme point. For other points on the R-E region boundary, the K-factor in each subband softly evolves between 0 and $\infty$ as we aim at lower rate and higher energy. Hence, in contrast to the diode linear model, we note that the diode nonlinearity does not only change the power allocation strategy across subbands but also the input distribution in each subband.

\par
\begin{figure}
   \centerline{\includegraphics[width=\columnwidth]{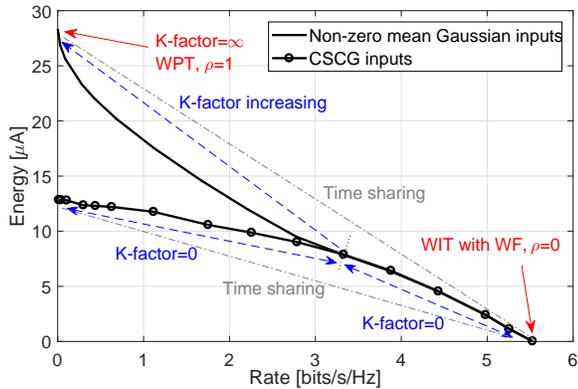}}
  \caption{R-E region for $N=16$ with a PS receiver and non-zero mean Gaussian inputs in multi-subband transmission \cite{Clerckx:2018b}. The average received power at the input the rectifier is fixed to -20 dBm and the SNR (defined as $P/\sigma^2$) is fixed to 20 dB in each subband. The rate has been normalized w.r.t. the bandwidth $N f_{\mathrm{w}}$. Hence, the x-axis refers to a per-subband rate. Note that the energy unit is $\mu$A because the energy metric used is $z_{\dc}$, which is a contribution to the output current.}
  \label{RE_region_nongaussian}
\end{figure}

 Fig. \ref{RE_region_nongaussian} illustrates the above discussion for a PS receiver and the significant enlargement of the R-E region by using non-zero mean inputs over CSCG, or equivalently by superposing a deterministic multisine waveform onto a modulated waveform (with CSCG symbols). This drastically contrasts with the conclusions obtained with the diode linear model. Recall indeed from Remark \ref{remark_linear} that, for the diode linear model, the input distribution does not impact the amount of harvested energy and there is no benefit in using a multisine waveform on top of the modulated waveform since both are equally suitable from an energy harvesting perspective. With the diode linear model, the use of non-zero mean inputs would have not provided any R-E region enhancement over the use of CSCG inputs. The R-E region enhancement in Fig. \ref{RE_region_nongaussian} also illustrates the gain obtained by accounting for the diode nonlinearity for SWIPT signal and system design. Here, again, $\mathcal{C}_{\R-\E}^\textnormal{L,TS}\subseteq\mathcal{C}_{\R-\E}^\textnormal{NL,TS}$, $\mathcal{C}_{\R-\E}^\textnormal{L,PS}\subseteq\mathcal{C}_{\R-\E}^\textnormal{NL,PS}$, $\mathcal{C}_{\R-\E}^\textnormal{L,Ideal}\subseteq\mathcal{C}_{\R-\E}^\textnormal{NL,Ideal}$.

\par Another interesting observation from Fig. \ref{RE_region_nongaussian} is the concavity-convexity of the R-E region boundary with non-zero mean inputs, which contrasts with the concavity of the region boundary for the CSCG inputs. We indeed note from Fig. \ref{RE_region_nongaussian} that the non-zero mean Gaussian inputs curve presents an inflection point, with the boundary being convex at low rate and concave at high rate. This has the consequence that TS can outperform PS for the diode nonlinear model, as illustrated in the figure. It is shown in \cite{Clerckx:2018b} that for a sufficiently large $N$ (e.g. $N=16$), TS is preferred at low SNR and PS at high SNR, but in general the largest convex hull is obtained by a combination of PS and TS.

\par The above discussion relies on Gaussian inputs. Moving one step closer to real-world digital communication system, we can leverage the diode nonlinear model and the above observations on SWIPT input distribution and waveform design to design SWIPT modulation based on finite constellations. In \cite{Bayguzina:2018}, the modulation of information symbols onto a multi-carrier energy-carrying waveform, resulting in a unified SWIPT modulated waveform, was studied. The authors adapted PSK
modulation to SWIPT requirements and showed the benefits of departing from conventional symmetric PSK modulation and adopt asymmetric PSK modulation, where all constellations points have the same magnitude but are uniformly distributed over $[-\delta,\delta]$ with $\delta\leq \pi$. By changing $\delta$ and optimizing the probability mass function of the constellation points, an asymmetric distribution of the constellation points was shown to enable a larger R-E region compared to that obtained with conventional symmetric PSK constellations.

\subsubsection{Multi-Antenna Transmission}
In a MISO setup $y_{n} =  \mathbf{h}_{n} \mathbf x_{n} +w_{n}$, it can be shown for general multi-band transmission that MRT in each subband is optimal \cite{Clerckx:2018b}. Hence, the optimal input symbol vector can be written as $\mathbf x_{n}=\bar{\mathbf{h}}_{n}^H x_{n}$ with $\bar{\mathbf{h}}_{n}=\mathbf{h}_{n}/\left\|\bar{\mathbf{h}}_{n}\right\|$ and $x_{n}$ designed according to the optimal input distribution of a SISO transmission for the diode nonlinear model.
\par The MIMO setup remains an interesting open problem. For multi-band MIMO transmission, one challenge is that the maximization of the energy at the $M_r$ receivers results in a coupled optimization of the frequency and the spatial domains, i.e., decoupling the spatial and frequency domain by first designing the spatial beamformer in each subband and then designing the power allocation across subbands is suboptimal contrary to the MISO case \cite{Huang:2017}.

\begin{observation}\label{observation_nonlinear} The use of the diode nonlinear model leads to four important observations. \textit{First}, the strategy that maximizes $P_{\rf}^r$ \textit{does not} maximize $P_{\dc}^r$. \textit{Second}, CSCG inputs cannot achieve the optimal R-E region boundaries. \textit{Third}, $\mathcal{C}_{\R-\E}^{\textnormal{NL,TS}}\not\subseteq \mathcal{C}_{\R-\E}^\textnormal{NL,PS}$. \textit{Fourth}, $\mathcal{C}_{\R-\E}^\textnormal{L,TS}\subseteq\mathcal{C}_{\R-\E}^\textnormal{NL,TS}$, $\mathcal{C}_{\R-\E}^\textnormal{L,PS}\subseteq\mathcal{C}_{\R-\E}^\textnormal{NL,PS}$, $\mathcal{C}_{\R-\E}^\textnormal{L,Ideal}\subseteq\mathcal{C}_{\R-\E}^\textnormal{NL,Ideal}$, i.e.\ the diode nonlinearity is beneficial to overall system performance.
\end{observation}

\par Comparing with Observation \ref{observation_linear}, we note that the diode nonlinear model has a deep impact on SWIPT design. It changes the input distribution and therefore the basic characteristics of the PHY and MAC layers such as modulation, waveform, spectrum use, and resource allocation as well as the receiver architecture. Though the beamforming designs for the diode linear and nonlinear models are identical for the point-to-point MISO system, namely both employ MRT, different designs are needed for MIMO systems. The importance of accounting for this diode nonlinearity in the designs and evaluations of WPT and SWIPT was first highlighted in \cite{Clerckx:2016b} and \cite{Clerckx:2018b}, respectively.

\subsection{Rate-Energy Tradeoff with The Saturation Nonlinear Model}\label{RE_sat}

In this subsection, we study the R-E tradeoff for the saturation nonlinear model from the resource allocation point of view, i.e., how the saturation nonlinearity of energy harvesting circuits affects resource allocation. To facilitate the presentation, we assume that the information signal is CSCG distributed and the use of an ideal receiver. This assumes that parameters $a$, $b$ and $P_{\mathrm{Sat}}$ of the saturation nonlinear model should be calculated for CSCG input signals\footnote{\label{Opt_sat_model}Since the saturation nonlinear model assumes a predefined waveform (with parameters $a$, $b$ and $P_{\mathrm{Sat}}$ fitted based on measurements obtained with that waveform), it may not be possible to explicitly define the capacity, as any change in the input distribution (and therefore waveform) would lead to changes in the model parameters $a$, $b$ and $P_{\mathrm{Sat}}$.
Nevertheless, what can be done is to model the saturation nonlinearity alternatively using output outage probability (OOP) constraints, which expresses the probability that the amplitude of the received signal outside a given interval is smaller than a threshold, as conducted in \cite{Varasteh:2018}. This enables to capture the saturation effect independently of the input signal. The capacity under \textit{average power constraints} and (OOP) constraints remains an open problem. For the low power range, CSCG is conjectured to be optimal, however for the higher delivery power range, CSCG is not. Nevertheless, it is unknown yet whether the optimal input distribution is made of an infinite number of mass points or finite or even, whether it is continuous. On the other hand, the capacity under \textit{average power and amplitude constraints} and (OOP) constraints is studied in \cite{Varasteh:2018}. It is shown that the amplitude of the optimal input is discrete with a finite number of mass points.}. First, we consider the case of single-subband transmission.

\subsubsection{Single-Subband Transmission} Let us write $x\sim\mathcal{CN}(0,P_{\rf}^t)$ with $P_{\rf}^t=\mathbb{E}\left[|x|^2\right]$. The optimal power allocation design for single-carrier transmission in SWIPT can be formulated as the following optimization problem
\begin{align}\label{eqn:optimization_prob_single_carrier}
\max_{P_{\rf}^t} \hspace{0.3cm}&  \log_2\left(1+\frac{\left|h\right|^2 P_{\rf}^t}{\sigma^2} \right) \\
\mathrm{subject\,\,to} \hspace{0.3cm} &P_{\rf}^t\leq P,\\
& P_{\dc}^r\geq \bar{E}.
\end{align}
The solution to the optimization problem in \eqref{eqn:optimization_prob_single_carrier} is trivial and the optimal power allocation is attained when $P_{\rf}^t=P$. Intuitively, for a given input signal distribution, the optimal strategy for maximizing the achievable rate with CSCG inputs and the total harvested power is to improve the received signal strength at the energy harvesting receiver as much as possible. This result actually aligns with that for the diode linear model and there is no tradeoff between rate and energy, i.e., the R-E region is again a rectangle similar to that of Fig. \ref{C_threereceivers}. However, the optimization problem based on the diode linear model in \eqref{SS_linear_opt}-\eqref{SS_linear_opt_EH_constraint} is always feasible for a sufficiently large maximum transmit power budget $P$. In contrast, if $\bar{E}>P_{\mathrm{Sat}}$ for the saturation nonlinear model, the problem becomes infeasible, even if $P\rightarrow \infty$.

\subsubsection{Multi-Subband Transmission}
Here, we study the power allocation problem for multi-carrier transmission. The optimal power allocation design can be formulated as the following optimization problem
\begin{align}\label{eqn:saturated_OFDM_problem_formulation}
\max_{\left\{P_0,...,P_{N-1}\right\}} \hspace{0.3cm}& \sum_{n=0}^{N-1} \log_2\left(1+\frac{\left|h_n\right|^2P_n}{\sigma^2} \right) \\
\mathrm{subject\,\,to} \hspace{0.3cm} &\sum_{n=0}^{N-1}P_n\leq P,\\
& \frac{\left[\frac{P_{\mathrm{Sat}}}{1+\exp\left(-a\left(\sum_{n=0}^{N-1}\left|h_n\right|^2P_n-b\right)\right)}
 - P_{\mathrm{Sat}}\Omega\right]}{1-\Omega}\geq \bar{E}, \label{eqn:auxiliary_non-linear _constraint_0}
\end{align}
where $P_n$ is the power allocated to subband $n$. In general,  \eqref{eqn:auxiliary_non-linear _constraint_0} is a convex constraint and the optimization problem in \eqref{eqn:saturated_OFDM_problem_formulation} can be solved efficiently via numerical convex program solvers.  However, in order to draw the connection between the problem formulations adopting the diode linear model and the saturation nonlinear model, we transform the optimization problem in \eqref{eqn:saturated_OFDM_problem_formulation} into the following equivalent problem

\begin{align}\label{eqn:saturated_OFDM_problem_formulation_equivalent}
\max_{\left\{P_0,...,P_{N-1},\beta\right\}} \hspace{0.3cm}& \sum_{n=0}^{N-1} \log_2\left(1+\frac{\left|h_n\right|^2P_n}{\sigma^2} \right) \\
\mathrm{subject\,\,to} \hspace{0.3cm} &\sum_{n=0}^{N-1}P_n\leq P,\\
 & \sum_{n=0}^{N-1}\left|h_n\right|^2P_n\geq \beta, \label{eqn:auxiliary_constraint}\\
& \frac{\left[\frac{P_{\mathrm{Sat}}}{1+\exp\left(-a\left(\beta-b\right)\right)}
 - P_{\mathrm{Sat}}\Omega\right]}{1-\Omega}\geq \bar{E}, \label{eqn:auxiliary_non-linear _constraint}
\end{align}
where $\beta$ is an auxiliary optimization variable representing the maximal received power at the receiver.
Note that the constraint in \eqref{eqn:auxiliary_constraint} is satisfied with equality at the optimal solution. By comparing the problem formulations in \eqref{eqn:saturated_OFDM_problem_formulation_equivalent} and \eqref{multicarrier_opt_problem}, both problems have almost identical structures, e.g. \eqref{multicarrier_opt_problem}-\eqref{harvested-energy constraint} versus \eqref{eqn:saturated_OFDM_problem_formulation_equivalent}-\eqref{eqn:auxiliary_constraint}, except that there is an extra constraint, i.e., \eqref{eqn:auxiliary_non-linear _constraint}. Therefore, similar to the case of the diode linear model, one would expect that there exists a non-trivial tradeoff between information transmission and energy transfer. Specifically, for the saturation nonlinear model, the amount of harvested DC power is maximized when the received power at the rectenna input is also maximized. Since \eqref{eqn:auxiliary_constraint} is an affine function with respect to $P_n$, the optimal power allocation to maximize the harvested DC power is to allocate $P$ to the subband with the best channel gain, i.e., $\max_{n\in\{0,\ldots,N-1\}} \left|h_n\right|$. Note that this observation is the same as the diode linear model. However, if the subbands are grouped into multiple chunks utilizing different energy harvesting circuits, then the power will be allocated over multiple chunks to avoid putting all the power to a chunk  where the corresponding energy harvesting circuits are already saturated.  On the other hand, to maximize the rate of the SWIPT system, standard WF solution can be adopted. Hence, \eqref{eqn:saturated_OFDM_problem_formulation_equivalent} can be solved by a modified WF solution similar to the one described in Section \ref{Linear_MS_section} for the diode linear model. Yet, the water level of the optimal power allocation for Problem \eqref{eqn:saturated_OFDM_problem_formulation_equivalent} is controlled by the dual variable associated with constraint \eqref{eqn:auxiliary_non-linear _constraint}, taking into account the saturation nonlinearity of the energy harvesting circuit. Furthermore, since the problem formulation can be transformed to an equivalent model using the diode linear model plus one additional constraint, the results of PS outperforming TS for the diode linear model should also hold for the saturation nonlinear model, though no works have been reported on the topic to verify such a claim.

\subsubsection{Multi-Antenna Transmission}
Consider a MIMO system with an ideal receiver for the saturation nonlinear model. The optimal resource allocation policy can be obtained by solving the following optimization problem
\begin{align}\label{eqn:TP_maximization}
\max_{\mathbf{Q}\in \mathbb{H}^{N_{\mathrm{T}}},\beta}\,\, &  \log_2\det\left(\mathbf{I}+\mathbf{H}\mathbf{Q}\mathbf{H}^H\right) \\
\mathrm{subject\,\,to}\hspace{0.3cm} &\Tr\left(\mathbf{Q}\right) \leq P,\\
 &  \Tr\left(\mathbf{H}\mathbf{Q}\mathbf{H}^H\right)\geq \beta, \label{eqn:auxiliary_constraint_MIMO}\\
& \frac{\left[\frac{P_{\mathrm{Sat}}}{1+\exp\left(-a\left(\beta-b\right)\right)}
 - P_{\mathrm{Sat}}\Omega\right]}{1-\Omega}\geq \bar{E}\label{eqn2:auxiliary_constraint_MIMO}.
\end{align}
The problem formulation in \eqref{eqn:TP_maximization}-\eqref{eqn2:auxiliary_constraint_MIMO} is similar to \eqref{MI_MIMO}-\eqref{MI_2_MIMO}, except that the auxiliary optimization variable $\beta$ is introduced in \eqref{eqn2:auxiliary_constraint_MIMO} taking into account the nonlinearity of the energy harvester for power allocation. Therefore, with a slight modification, the generalized multiple eigenmode transmission of subsection \ref{Linear_MA_section}, introduced in \cite{Zhang:2013}, remains optimal. Besides, the tradeoff between rate and energy for the saturation nonlinear model is similar to that for the diode linear model.

\begin{observation}\label{observation_saturation}
The use of the saturation nonlinear model leads to four important observations. \textit{First}, the optimization problem adopting the saturation nonlinear model can be transformed into an equivalent optimization problem adopting the diode linear model with one additional constraint, e.g. equation \eqref{eqn2:auxiliary_constraint_MIMO}. This implies that the saturation nonlinearity is detrimental to overall system performance. \textit{Second}, the strategy that maximizes $P_{\rf}^r$ also maximizes $P_{\dc}^r$ (similarly to the diode linear model). \textit{Third}, PS is expected to outperform TS (similarly to the diode linear model). \textit{Fourth}, as a consequence of footnote 8, CSCG inputs cannot in general achieve the optimal R-E region boundaries in the presence of the saturation nonlinearity (similarly to the diode nonlinear model).
\end{observation}

\subsection{Extension and Future Work}

The above discussions highlight how significantly the signal design depends on the underlying energy harvester model. Several interesting research avenues arise.
\par First, in Sections \ref{L_section}, \ref{NL_section}, and \ref{RE_sat} and the related literature, perfect CSIT and CSIR are assumed. Acquiring CSIT is a challenge due to the limited energy available at the terminals. To that end, various CSI acquisition techniques have been developed for WPT and SWIPT assuming the diode linear model in \cite{Zeng:2017,Xu_Zhang:2014,Zeng_Zhang:2015,Xu_Zhang:2015a,Xu_Zhang:2016,Abeywickrama:2018} and \cite{Liu_Quek:2015}, respectively. However, little is known about how to design CSI acquisition strategies for the nonlinear models. A promising attempt was made in \cite{Huang:2018} where codebooks of waveform precoders (spanning jointly the space and frequency domains) were designed for the diode nonlinear model using a framework reminiscent of the generalized Lloyd’s algorithm. It was shown that the diode nonlinear model-based waveform design with limited feedback outperforms the diode linear model-based waveform design relying on perfect CSIT. This also leads to interesting new challenges for CSIT acquisition for SWIPT with the nonlinear energy harvester models.
\par Second, the diode nonlinearity leads to a re-thinking of the optimal input distribution, modulation, and waveform. SWIPT with practical modulations based on finite constellations has been studied in \cite{Zewde:2015,Kim_Kim:2016,Zhu_Xiao:2017,Kim_Choi:2018} for the diode linear model. The design of practical and efficient modulations and waveforms for SWIPT with nonlinear energy harvester models remains virtually untouched. As a consequence of the diode nonlinearity, asymmetric PSK constellations have appeared in \cite{Bayguzina:2018} and were shown to outperform conventional symmetric PSK constellations. Nevertheless the extension to more general non-constant modulus modulations should provide additional performance benefits. Moreover, considering the benefits of non-zero inputs in multi-subband transmission, constellations with a non-zero offset are also an attractive option. In all cases, the shaping of the complex constellation points will have to be revisited and optimized in order to maximize some R-E or error probability-energy metrics. In general, as the required energy $\bar{E}$ increases, the optimal design of the constellation would shift away from the classical QAM design.
\par Third, changes in modulation design for SWIPT would also lead to some changes in error-correcting code design. Hence, coding for optimized SWIPT constellations will need to be re-visited. This should not be confused with \cite{Fouladgar:2014,Tandon:2016} that are motivated by the underflow/overflow of batteries, not by the diode nonlinearity. WIPT design for short packets and finite length coding is also of interest, though currently its analysis has been limited to the diode linear model \cite{KHP17}.
\par Fourth, characterizing the optimal input distributions for the diode nonlinear model and the saturation nonlinear model remains largely open problems. Though some works have recently appeared in \cite{Varasteh:2017,Morsi:2017,Varasteh:2017b,Varasteh:2018}, efforts have been limited to the single-subband single-antenna settings. Extensions to multi-subband and multi-antenna (MIMO) settings remain completely open problems. Note that in such setups, both the input distribution and the power allocation across subbands/eigenmodes will differ from conventional CSCG and WF strategies used for the diode linear model.
\par Fifth, the design of secure SWIPT will need to be revisited in light of the nonlinearity. Designs of secure SWIPT have appeared in \cite{Liu_Chua:2014,Ng_Schober:2014,Wu_Zhong:2016,Chen_Liu:2016,Liu_Popovski:2016,Xing:2015,Nasir:2017} and \cite{JR:non-linear_EH_HWI,Niu_Zhang:2017} for the linear diode model and the saturation nonlinear model, respectively. However, no work exists on secure SWIPT for the diode nonlinear model.
\par Sixth, the signal design should also be re-visited for WPCN and WPBC since the nonlinearity will have significant impact on the modulation, waveform, and resource allocation designs for those systems as well. For instance, WPBC was considered in \cite{Clerckx:2017b} and it was shown that multisine waveforms can be designed to account for the diode nonlinearity to enhance the SNR at the reader and the harvested energy at the tag. An SNR-energy tradeoff exists in WPBC because SNR maximization at the reader and energy maximization at the tag do not lead to the same waveform design and power allocation strategy.
\par Finally, it would be worth connecting the above findings and advances to other fields and applications subject to nonlinearity such as intermodulation distortion, optical channels, magnetic recording, PA saturation on OFDM. In optical communications (and other applications) the nonlinearity is commonly compensated and transmission is performed using constellations approximating the zero-mean Gaussian distribution optimum for AWGN channels (e.g. ring constellations) \cite{Essiambre:2010}. The information theoretic limits of optical channels are studied by modelling the nonlinear optical communication channel as a linear channel with a multiplicative noise or using a finite-memory model with additive noise \cite{Essiambre:2010,Agrell:2014}. On the contrary, in SWIPT, the diode nonlinearity is exploited in the signal design and in the characterization of the R-E region, therefore leading to non-zero mean Gaussian inputs and enlarged region compared to that obtained with zero-mean inputs.

\section{Multi-User WIPT}\label{multi_user_WIPT_section}
\begin{figure}[t]
 \centering
\includegraphics[width=3.5 in]{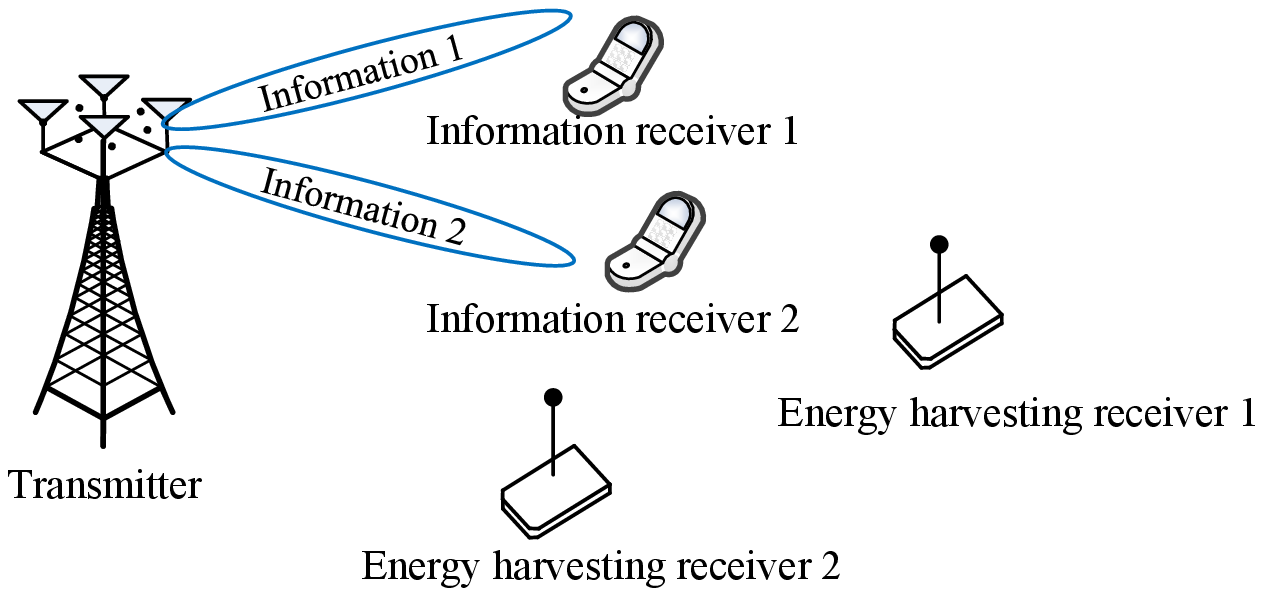}
 \caption{A multi-user SWIPT downlink model with a transmitter,  $K=2$ information receivers, and $J=2$ energy harvesting receivers.  } \label{SWIPT_multiuser_model_Kwan}
\end{figure}

WIPT systems exploit the broadcast nature of wireless channels, which opens the possibility of one-to-many charging. Hence, it is important to study the impact of linear and nonlinear energy harvesting from a multi-user perspective. With this in mind, in this section, we consider a downlink multi-user SWIPT system consisting of a transmitter equipped with $M_\mathrm{t}$ antennas as well as $K$ single-antenna IRs and $J$ single-antenna ERs, as illustrated in Fig. \ref{SWIPT_multiuser_model_Kwan}. This setup is also denoted as ``SWIPT with separated information and energy receivers'' in Fig. \ref{SWIPT_figure}. In contrast to single-user WIPT, in multi-user SWIPT, how to deal with the co-channel interference due to the simultaneous transmissions to multiple users is a critical issue. Unlike traditional wireless communications where the interference is treated as an undesired phenomenon for WIT, it can be exploited for wireless energy harvesting \cite{Park:2013,Liu_Chua:2013}, which reveals an interesting new research on interference management in WIPT systems.

\subsection{Rate-Energy Tradeoff with The Linear Model}

We start with the diode linear model in the above multi-user SWIPT system. Under this setup, the transmitter broadcasts a signal $\mathbf{x}\in \mathbb{C}^{M_\mathrm{t} \times 1}$ to all the users. Generally speaking, $\mathbf{x}$ is comprised of $K$ information beams (one for each IR) and $J$ energy beams (one for each ER), i.e.,
\begin{align}\label{eqn:transmit signal multiuser}
\mathbf{x}=\sum\limits_{k=1}^K\mathbf{p}_k s^{{\ID}}_k+\sum\limits_{j=1}^J\mathbf{v}_j s^{{\EH}}_j,
\end{align}
where $s^{{\ID}}_k\in \mathcal{CN}(0,1)$ and $\mathbf{p}_k\in \mathbb{C}^{M_\mathrm{T}\times 1}$ denote the information-bearing signal and the corresponding energy beamforming vector for the $k$th IR, respectively, and $s^{{\rm EH}}_j\in \mathcal{CN}(0,1)$ and $\mathbf{v}_j\in \mathbb{C}^{M_\mathrm{t}\times 1}$ denote the energy-carrying signal and the corresponding energy beamforming vector for the $j$th ER, respectively. Then, the received signals at the $k$th IR and the $j$th ER are respectively expressed as
\begin{align}
y^{{\ID}}_k & =\mathbf{h}_k\mathbf{x}+w^{{\ID}}_k \nonumber \\
& =\mathbf{h}_k\sum\limits_{i=1}^K\mathbf{p}_i s^{{\ID}}_i+\mathbf{h}_k\sum\limits_{j=1}^J\mathbf{v}_j s^{{\EH}}_j+w^{{\ID}}_k, ~ \forall k, \label{eqn:received_signal_information_receiver}\\
y^{{\EH}}_j & =\mathbf{g}_j\mathbf{x}+w^{{\EH}}_j \nonumber \\
& =\mathbf{g}_j\sum\limits_{k=1}^K\mathbf{p}_k s^{{\ID}}_k+\mathbf{g}_j\sum\limits_{i=1}^J\mathbf{v}_i s^{{\EH}}_i+w^{{\EH}}_j, ~ \forall j,  \label{eqn:received_signal_energy_receiver}
\end{align}
where $\mathbf{h}_k\in \mathbb{C}^{1\times M_\mathrm{t}}$ and $\mathbf{g}_j\in \mathbb{C}^{1\times M_\mathrm{t}}$ denote the channels from the transmitter to the $k$th IR and the $j$th ER, respectively, and $w^{{\ID}}_k\sim \mathcal{CN}(0,\sigma^2)$ and $w^{{\EH}}_j\sim \mathcal{CN}(0,\sigma^2)$ denote the Gaussian noise at the $k$th IR and $j$th ER, respectively. We assume for simplicity that the noise powers $\sigma^2$ are identical at all receivers. It is also assumed that the noise does not contribute to the harvested energy.

\par It is worth noting that the information-bearing signals $s^{{\ID}}_k$'s must be random, but the energy signals $s^{{\EH}}_j$'s can be pseudo-random since they do not contain any information\footnote{This is a consequence of the diode linear model as explained in Remark \ref{remark_linear}. For the diode nonlinear model, the choice of the energy signal, modulated or deterministic and its distribution, would have an influence on the ultimate performance, similarly to the single-user SWIPT in Section \ref{NL_section}.}. As a result, it is theoretically possible to cancel the interference caused by the energy signals\footnote{Recall that such an interference cancellation of the energy signal was also used in Section \ref{MS_section_NL_SU} for single-user SWIPT.} if they are pre-stored at the IR side. Reference \cite{Xu:2014} studies both the cases that the interference caused by the energy signals can or cannot be canceled by the information receivers. Interestingly, it is shown in \cite{Xu:2014} that to achieve the optimal R-E tradeoff, no dedicated energy signals should be used in the case that the energy signals cannot be canceled by the IRs, i.e., $\mathbf{v}_j=\mathbf{0}$, $\forall j$; while no more than one energy signal is sufficient in the other case that the energy signals can be canceled by the IRs\footnote{Note that this is again a consequence of the diode linear model. For the diode nonlinear model, it was indeed shown in \cite{Clerckx:2018b} that, in the event where the energy signal is not eliminated (and therefore treated as noise) by the communication receiver, the energy signal is still useful and does help enlarging the R-E region.}.

\par In the rest of this section, we mainly focus on the case when the interference caused by the energy signals cannot be canceled by the IRs. In this case, since $\mathbf{v}_j=\mathbf{0}$, $\forall j$, under the optimal solution \cite{Xu:2014}, the received signals at the information and energy receivers given in (\ref{eqn:received_signal_information_receiver}) and (\ref{eqn:received_signal_energy_receiver}) respectively reduce to
\begin{align}
& y^{{\ID}}_k=\mathbf{h}_k\sum\limits_{i=1}^K\mathbf{p}_is^{{\ID}}_i+w^{{\ID}}_k, ~ \forall k, \label{eqn:received_signal_information_receiver_1}\\
& y^{{\EH}}_j=\mathbf{g}_j\sum\limits_{k=1}^K\mathbf{p}_ks^{{\ID}}_k+w^{{\EH}}_j, ~ \forall j.  \label{eqn:received_signal_energy_receiver_1}
\end{align}

\par Under the above model, for IR $k$, its signal-to-interference-plus-noise ratio (SINR) to decode the message $s^{{\ID}}_k$ is
\begin{align}\label{eqn:SINR multiuser}
\gamma_k=\frac{\mathbf{p}^H_k\mathbf{H}_k\mathbf{p}_k}
{\sum_{j\neq k} \mathbf{p}_j^H\mathbf{H}_k\mathbf{p}_j+\sigma^2}, ~ \forall k,
\end{align}
where $\mathbf{H}_k=\mathbf{h}_k^H\mathbf{h}_k$. Moreover, for ER $j$, its harvested DC power for the diode linear model is proportional to $\sum_{k=1}^K\|\mathbf{g}_j\mathbf{p}_k\|^2$, $\forall j$.

\par To achieve the optimal R-E tradeoff, the information beams $\mathbf{p}_k$'s can be jointly optimized to maximize the sum-energy harvested by the $J$ ERs subject to the transmit power constraint as well as each IR's SINR constraint, i.e.,

\begin{align}\label{eqn:sum-energy maximization multiuser}
\max_{\{\mathbf{p}_k\}} \hspace{0.3cm}& \sum\limits_{j=1}^J \sum\limits_{k=1}^K\|\mathbf{g}_j\mathbf{p}_k\|^2 \\
\textnormal{subject to} \hspace{0.3cm} & \sum\limits_{k=1}^K\|\mathbf{p}_k\|^2 \leq P, \label{eqn:sumpower_constraint}\\
& \frac{\mathbf{p}^H_k\mathbf{H}_k\mathbf{p}_k}
{\sum_{j\neq k} \mathbf{p}_j^H\mathbf{H}_k\mathbf{p}_j+\sigma^2}\geq \bar{\Gamma}_k, ~~~ \forall k, \label{eqn:SINR constraint}
\end{align}
where $P$ denotes the total power available at the transmitter, and $\bar{\Gamma}_k$ denotes the SINR target of the $k$th IR.

\par The optimal beamforming solution to the above problem can be obtained by either the semidefinite relaxation (SDR) technique or the uplink-downlink duality technique, as further detailed in \cite{Xu:2014}. As an alternative beamforming design, \cite{Son:2014} investigates a more practical design where the beamforming vectors are initialized using the well established zero-forcing beamforming (ZFBF) and a simple algorithm is proposed to successively update the beamformers so as to maximize the total harvested energy subject to SINR constraints at the IRs.

\par Besides the above setup, there are other studies under the general broadcast SWIPT model with separated information/energy receivers. For example, the capacity region of the IRs subject to each ER's energy harvesting constraint is characterized in \cite{Luo:2015}, in which the interference caused by the energy signals is assumed to be perfectly canceled by the IRs, while the transmitter uses the optimal dirty paper coding-based non-linear precoding strategy for information transmission. Moreover, the linear beamforming design for achieving the optimal tradeoff between the secrecy rate of the IRs and energy harvested by the ERs is studied in \cite{Liu_Chua:2014} and \cite{Ng_Schober:2014}, in which the ERs are treated as potential eavesdroppers. Precoder designs for the general multi-user MIMO SWIPT with multiple antennas at the transmitter, IRs and ERs have been studied in \cite{Song:2016}. In contrast to previous precoder designs that focused on maximizing the information rate, \cite{Song:2016} derives a simple solution using a weighted minimum mean squared error (WMMSE) criterion. Finally, robust beamforming under the
assumption of imperfect CSIT to maximize the worst-case harvested energy at the ER while guaranteeing a target rate at the IR has been studied in \cite{Xiang:2012}.

\par Furthermore, it is worth noting that in addition to the case of separated information/energy receivers, various other multi-user SWIPT settings are also studied in the literature. For example, \cite{Shi:2014} investigates a multi-user broadcast model in which each user adopts the PS strategy for splitting a portion of its received signal for information decoding and the remaining portion for energy harvesting, i.e., the so-called co-located information and energy receiver. In this setting, the transmit beamforming and the receive PS ratios are jointly designed to achieve the optimal R-E tradeoff. Further, an OFDM-based multi-user broadcast SWIPT system is considered in \cite{Zhou:2014}, in which the subchannel allocation, as well as the transmit power and receive PS ratio at each subchannel are jointly optimized to achieve the best R-E tradeoff. It was shown that PS always outperforms TS in the general multi-user multi-subband SWIPT transmissions.

\par Moveover, the multi-user SWIPT system is also studied in the interference channel setting, where multiple transmitters send independent messages to their corresponding receivers, and at the same time cooperatively transmit power wirelessly to the receivers. Specifically, \cite{Park:2013} and \cite{Park:2014} consider a multi-user SWIPT system under the MIMO interference channel setup with TS receivers, where the interference caused by the energy signals is assumed unknown at the information receivers and thus cannot be canceled. Given that the TS strategy is adopted by each receiver (either in information decoding or in energy harvesting mode at any time instant), the precoding designs of all transmitters are jointly optimized in terms of R-E tradeoff. In contrast, \cite{Lee:2015} exploits the fact that despite the lack of coordination between the transmitters for coordinated information transmission, energy beamforming can be performed across all the transmitters since the energy signals are pseudo-random and thus can be pre-stored at all the transmitters as well as all the receivers for interference cancellation. Under this setup, the joint optimization of transmit precoding and receiver TS strategy is revisited in \cite{Lee:2015}, where a new transmitter-side PS approach is proposed. Some subsequent works on precoder optimization for SWIPT multi-antenna interference channel have appeared in \cite{Park:2015,HLee:2015,Timotheou:2014,Zong:2016,Shi:2014a}, with also additional considerations for limited feedback \cite{Park:2015}.

\par Other important multi-user scenarios include multicasting \cite{Khandaker:2014} and multiple access channel \cite{Fouladgar:2012,Amor:2017}. Furthermore, SWIPT systems are also investigated in multi-user cooperative communications under various different setups, such as with TS- and/or PS-based half-duplex relaying \cite{Ding:2014,Nasir:2013,Huang:2015}, as well as full-duplex relays with simultaneous information transmission and energy harvesting \cite{Zhong:2014,Zeng:2015}. Other relaying setups include SWIPT in relay system with multiple antennas at all nodes \cite{Huang:2016r}, relay interference channels \cite{He_Chen:2015}, interference-aided energy harvesting relay \cite{Gu:2015} and relay selection \cite{MSS15}.

\par Finally, stochastic geometry has been used to analyze the performance of various large-scale SWIPT networks in microwave and millimeter-wave bands \cite{Huang_Lau:2014,Krikidis:2014a,KAH16,Akbar:2016,Di_Renzo:2017,Park:2018}.

\begin{observation}\label{observation_MU_linear} The observations made for the diode linear model in single-user SWIPT carry over to the multi-user SWIPT. \textit{First}, the strategy that maximizes the total received RF power (across all users) maximizes the total harvested DC power. \textit{Second}, CSCG inputs for the information-bearing signal and (pseudo-random) CSCG inputs for the energy-bearing signal, if needed, are sufficient to achieve the R-E region boundaries. \textit{Third}, PS outperforms TS.
\end{observation}

\subsection{Rate-Energy Tradeoff with The Nonlinear Models}

In this subsection, we study the multi-user SWIPT system described by (\ref{eqn:received_signal_information_receiver_1}) and (\ref{eqn:received_signal_energy_receiver_1}), but with a nonlinear energy harvesting model. In this case, the SINR for decoding $s^{{\rm ID}}_k$ at the $k$th IR is still expressed as (\ref{eqn:SINR multiuser}). However, by adopting the saturation nonlinear model, the total harvested power at the $j$th ER is given by
\begin{eqnarray}\label{eqn:EH_non_linear_multiuser}
 \hspace*{-5mm}P_{\mathrm{dc}_j}^r\hspace*{-
 2mm}&=&\hspace*{-2mm}
 \frac{[\Psi_{\mathrm{dc}_j}
 \hspace*{-0.5mm}- \hspace*{-0.5mm}P_{\mathrm{Sat}_j}\Omega_j]}{1-\Omega_j},\, \Omega_j=\frac{1}{1+\exp(a_jb_j)},\\
 \hspace*{-5mm}\Psi_{\mathrm{dc}_j}\hspace*{-2mm}&=&\hspace*{-2mm} \frac{P_{\mathrm{Sat}_j}}{1+\exp\Big(\hspace*{-0.5mm}-a_j(P_{\rf_j}^r-\hspace*{-0.5mm}b_j)\Big)},\,\mbox{and}\\
P_{\rf_j}^r&=&\sum_{k=1}^K\Tr\Big(\mathbf{p}_k\mathbf{p}^H_k\mathbf{g}_j^H\mathbf{g}_j\Big)
  \end{eqnarray}
is the received RF power at ER $j$.

\par The  sum-power maximization problem given in (\ref{eqn:sum-energy maximization multiuser}) for the diode linear model is thus modified to
\begin{align}\label{eqn:Energy_maximization_saturated_model_multiuser}
\max_{\{\mathbf{p}_k\}} \hspace{0.3cm}& \sum\limits_{j=1}^J P_{\mathrm{dc}_j}^r \\
\textnormal{subject to} \hspace{0.3cm} & (\ref{eqn:sumpower_constraint}), ~ (\ref{eqn:SINR constraint}).
\end{align}

\par Note that it is challenging to solve the optimization problem in \eqref{eqn:Energy_maximization_saturated_model_multiuser} since the objective function is in the form of sum-of-ratios. In \cite{Boshkovska:2015}, the authors have proposed a series of transformations to transform the objective function into an equivalent objective function in subtractive form, which enables the design of an efficient iterative optimal resource allocation algorithm. In each iteration, a rank-constrained semidefinite program (SDP) is solved optimally by SDP relaxation. Note that the optimal solution to the problem in \eqref{eqn:Energy_maximization_saturated_model_multiuser} is beamforming.  However, the optimal beamforming solution structure for the diode linear model is different from that of the saturation nonlinear model. In an extreme case, when the channels of the $J$ energy harvesting receivers are orthogonal to each other, i.e., $\mathbf{g}_i\mathbf{g}_j^H=0,\,\forall i\neq j$, and $\bar{\Gamma}_k=0$, the optimal beamforming for the diode linear model will perform MRT and allocate all transmit power in the direction of  $\max_{j\in\{1,\ldots,J\}} \left\|\mathbf{g}_j\right\|$. However, for the saturation nonlinear model, the optimal beamforming design is the transmission via the maximum eigenmode of matrix $\sum_{j=1}^J \beta_j\mathbf{g}_j^H\mathbf{g}_j$. In particular, $\beta_j\ge 0$ are dual variables related to the constraints of the received power at each ER, cf. eq (10) in \cite{Boshkovska:2015},  which act as weights for determining the beamforming direction. In fact, the value of $\beta_j$ becomes smaller  when the $j$th ER enters the saturation region. In other words, the dual variables prevent the transmitter from allocating  exceedingly large powers in the directions of receivers whose energy harvesting circuits are already saturated.

 \begin{figure}[t]
 \centering
\includegraphics[width=3.5 in]{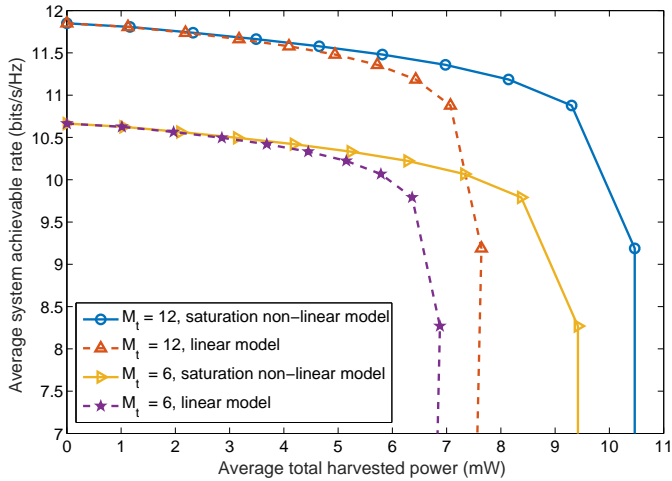}
 \caption{A comparison of R-E region achieved by resource allocation algorithms designed based on the saturation nonlinear model and the diode linear model, respectively.} \label{fig:saturated_model_simulation}
\end{figure}

\par In Fig. \ref{fig:saturated_model_simulation}, we show the R-E tradeoff region of the considered downlink multi-user SWIPT system. We assume that there are $K=1$ IR and $J=5$ ERs. We adopt the same simulation parameters as in \cite{Boshkovska:2015}. For comparison, we also show the performance of a baseline scheme in Fig. \ref{fig:saturated_model_simulation}. For the baseline scheme, the resource allocation algorithm is optimized for maximization of the total system harvested power according to the diode linear model subject to constraints (\ref{eqn:sumpower_constraint}), (\ref{eqn:SINR constraint}). Then, the baseline scheme is applied for resource allocation in the SWIPT system with the saturation nonlinear model. As can be observed, the baseline scheme designed for the diode linear model can only achieve a strictly smaller R-E region due to the resource allocation mismatch. In fact, the baseline scheme does not utilize the system resources efficiently since it causes saturation at some ERs and under-utilization at others. Also, it can be observed that increasing the number of transmit antennas can increase the R-E region significantly. This is because additional transmit antennas equipped at the transmitter provide extra spatial degrees of freedom which facilitate a more flexible resource allocation.

\par When it comes to the diode nonlinear model, no works currently exist on multi-user SWIPT though it is expected that all observations made in single-user SWIPT do carry over to the multi-user SWIPT. In particular, reference \cite{Varasteh:2017b}, though based on a point-to-point system model with an ideal receiver, is actually also applicable to a scenario with two separate receivers, namely one IR and one ER.

\begin{observation}\label{observation_MU_nonlinear}
The observations made for the \textit{saturation nonlinear model} in single-user SWIPT do not all carry over to the multi-user SWIPT. Indeed, the strategy that maximizes the total received RF power (across all users) does not maximize the total harvested DC power (in contrast to the single-user case and to the diode linear model). Depending on the system operation regime, the beamforming direction is steered generally towards a different direction compared to the problem formulation adopting the diode linear model. The observations made for the \textit{diode nonlinear model} in single-user SWIPT are expected to carry over to multi-user SWIPT.
\end{observation}

\subsection{Extension and Future Work}

The use of the diode linear model in multi-user SWIPT design has been extensively studied. Research is on the other hand at its infancy when it comes to multi-user SWIPT design for the nonlinear models. A number of promising research avenues are discussed below.
\par First, no works exist on multi-user SWIPT for the diode nonlinear model. Similarly to the point-to-point case, the diode nonlinear model will also lead to new input distribution, modulation, and waveform designs in the multi-user SWIPT setup. A first interesting avenue would be the design of multi-user SWIPT waveforms for the broadcast and interference channels. To that end, a good starting point might be the multi-user WPT waveform optimization framework in \cite{Huang:2017} and the superposed SWIPT waveforms of \cite{Clerckx:2018b} so as to design and optimize multi-user SWIPT waveforms. The benefits of non-zero Gaussian inputs in multi-user SWIPT systems could then be assessed. Another interesting research avenue is the study of the fundamental limits of broadcast and interference channels for the diode nonlinear models so as to extend the results of \cite{Varasteh:2017,Varasteh:2017b} to multi-user communications.
\par Second, in view of the significant changes brought by nonlinearity, it is of interest to re-think the SWIPT architectures for broadcast, multiple access, interference, and relay channels with and without secrecy constraints. The performance analysis of large SWIPT networks with nonlinear energy harvester models is also of interest.
\par Third, the diode nonlinearity is expected to have significant impacts on other forms of multi-user WIPT such as WPCN and WPBC. A recent work in \cite{Zawawi:2018} has investigated the impact of the diode nonlinearity on multi-user waveform design for WPBC. In contrast to point-to-point WPBC, multi-user WPBC is subject to multi-user interference and the transmit waveform needs to be optimized so as to maximize the SINR at the reader and the energy harvested at each tag, while exploiting the benefits of the diode nonlinearity, the channel frequency diversity gain, and the multi-user diversity gain.
\par Fourth, the multi-user system SWIPT model discussed above, as per \eqref{eqn:SINR multiuser}, assumes linearly-precoded multi-user transmission with any residual multi-user interference fully treated as noise. A more general and powerful transmission framework would consist in partially decoding interference and partially treating interference as noise through rate-splitting \cite{Clerckx:2016a}. Such a rate-splitting strategy has been shown to outperform conventional linear precoding in a wide range of network loads (underloaded and overloaded regimes) and user deployments (with a diversity of channel directions, channel strengths and qualities of Channel State Information at the Transmitter)\cite{Joudeh:2016,Mao:2018}. The use of rate-splitting for multi-user SWIPT for both the linear and nonlinear energy harvester models remains an uncharted research area.

\section{Prototyping, Experimentation, and Validation}\label{proto_section}

\par Demonstrating the feasibility and validating the aforementioned signal theory and design through prototyping and experimentation remains a largely open challenge. It requires the implementation of a closed-loop WPT/WIPT architecture with a real-time over-the-air transmission based on a frame structure switching between a channel acquisition phase and wireless power and information transfer phase. The channel acquisition needs to be done at the millisecond level (similarly to CSI acquisition in communication). Different blocks need to be built, namely channel estimation, modulation, channel-adaptive waveform and beamforming, rectenna and SWIPT receiver. The first prototype and early results of a closed-loop WPT architecture based on dynamic channel acquisition were reported in \cite{Kim:2017}, with further enhancements in \cite{Kim:2018}, for channel-adaptive waveform and modulation optimization and in \cite{ChoiKim2017a,ChoiKim2018a} for channel-adaptive beamforming optimization, conducted at Imperial College London and Sungkyunkwan University, respectively. Importantly, the channel acquisition needs very low circuit power consumption at the receiver. This is because the net energy harvested at a sensor node should be sufficient to sustain its energy neutral operation, as demonstrated in \cite{ChoiKim2017a,ChoiKim2018a}.

\par In the sequel, we illustrate some experimental results in the low-power regime and show that they validate the diode nonlinear model-based signal theory and design. We then discuss the use of multi-antenna beamforming to further increase the harvested DC power.

\subsection{Single-Subband Transmission}\label{SS_exp}

\begin{figure}
   \centerline{\includegraphics[width=\columnwidth]{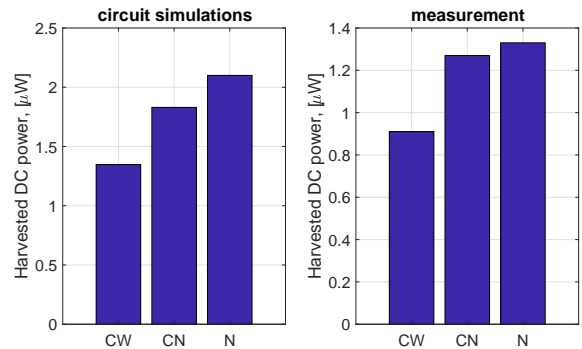}}
  \caption{Effect of input distribution in single-subband transmission on harvested DC power \cite{Clerckx:2018b,Kim:2018} (CW refers to continuous wave, CN to CSCG input, and N to real Gaussian input).}
  \label{modulation}
\end{figure}

Using the circuit simulator of \cite{Clerckx:2016b,Clerckx:2018b,Clerckx:2017} and the prototype of \cite{Kim:2017,Kim:2018}, Fig. \ref{modulation} illustrates circuit simulations and experimentation of the amount of harvested energy using three different input distribution when the average received power at the input of the rectenna is $P_{\rf}^r=-20$ dBm: a continuous wave (CW) with average input power $P_{\rf}^r$, a CSCG (CN) input $\sim\mathcal{CN}(0,P_{\rf}^r)$ and a real Gaussian (N) $\sim\mathcal{N}(0,P_{\rf}^r)$. We note that the circuit simulations and the experimentation both show that $P_{\dc,\mathrm{N}}\geq P_{\dc,\mathrm{CN}}\geq P_{\dc,\mathrm{CW}}$, namely that a higher DC power can be harvested from a real Gaussian input compared to a CSCG input and a CW. This confirms the conclusions drawn from the theoretical analysis of the diode nonlinear model in Section \ref{NL_section}. Moreover, recall that according to the linear diode model, a continuous wave, a CSCG and a real Gaussian with the same average RF power $P_{\rf}^r$ should yield the same DC power $P_{\dc}^r$ at the output of the rectifier. Clearly, this is not the case from Fig. \ref{modulation}. Hence, those simulations and measurements also invalidate Remark \ref{remark_linear} and the second observation in Observation \ref{observation_linear} for the diode linear model. Other recent measurement campaigns studying the effect of conventional QAM and PSK modulation on harvested energy and data rate have appeared in \cite{Claessens:2017a}.

\subsection{Multi-Subband Transmission}\label{MS_exp}

\begin{figure}
   \centerline{\includegraphics[width=\columnwidth]{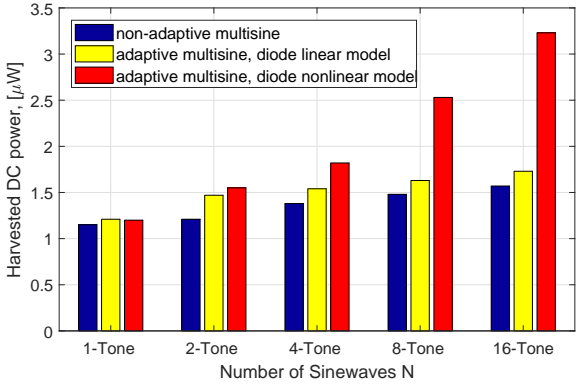}}
  \caption{Measurement of harvested DC power of multisine waveform \cite{Kim:2018}.}
  \label{waveform}
\end{figure}

Making use of the prototype in \cite{Kim:2017,Kim:2018}, Fig. \ref{waveform} illustrates the measured DC power levels at the output of the rectenna with three different transmit multisine waveforms with $N$ sinewaves uniformly spaced within a 10 MHz bandwidth ($\Delta_f=B/N$ with $B=10$MHz). A SISO setup is considered with the transmitter and receiver separated by about 5 m. The transmit power was set to 33 dBm and measured RF power at the input of the rectenna varied from -18 to -25 dBm. A non-adaptive in-phase multisine with uniform power allocation and two channel-adaptive multisine waveforms designed based on the diode linear and nonlinear models are adopted. The diode linear model-based design allocates all transmit power to the sinewave corresponding to the largest frequency-domain channel so as to maximize $e_2$ (see Section \ref{Linear_MS_section}). Doing so, the diode linear model-based design maximizes the input power $P_{\rf}^{r}$ to the rectifier. On the other hand, the diode nonlinear model-based design allocates power non-uniformly to all sinewaves (see Section \ref{NL_section}) so as to benefit from the diode nonlinearity and the channel frequency diversity to maximize $e_2 e_3$ \cite{Clerckx:2016b}. The diode nonlinear model-based design does not maximize $P_{\rf}^{r}$, but rather maximizes $P_{\dc}^{r}$ accounting for the rectifier nonlinearity. Hence $P_{\rf}^{r}$ achieved by the diode nonlinear model-based design is lower than that obtained with the diode linear model-based design. Nevertheless, comparing the two adaptive waveforms in Fig. \ref{waveform}, we note the diode nonlinear model-based design leads to significantly larger output DC power $P_{\dc}^{r}$ than the diode linear model-based design. We also note that the channel-adaptive waveform provides significant gains over non-adaptive designs if the diode nonlinearity is properly accounted for in the waveform design. These measurements confirm the importance and the benefits of accounting for the diode nonlinearity in WPT/WIPT system design and validate the theoretical analysis for the diode nonlinear model in Section \ref{NL_section} and \cite{Clerckx:2016b,Clerckx:2017}. In particular, measurements validate the first observation of Observation \ref{observation_nonlinear}, namely that the strategy that maximizes $P_{\rf}^r$ does not necessarily maximize $P_{\dc}^r$. This also invalidates the first observation of Observation \ref{observation_linear}. Finally, it is also important to note that multipath and channel frequency selectivity was also shown using theoretical analysis and circuit simulations in \cite{Clerckx:2016b} to have a significant impact on waveform design and harvested energy. Measurements in \cite{Kim:2017,Kim:2018} and \cite{Pan:2017} have independently confirmed those observations.

\subsection{Multi-Antenna Transmission}\label{MA_exp}

\begin{figure}
   \centerline{\includegraphics[width=\columnwidth]{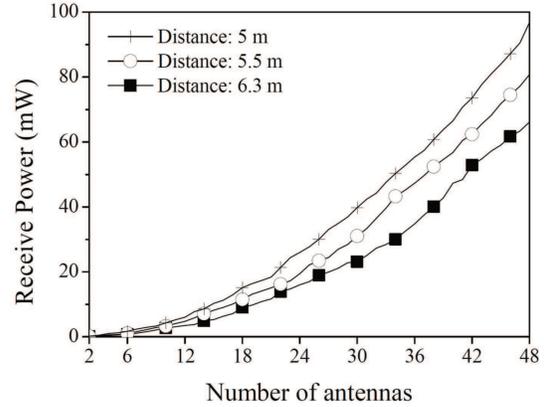}}
  \caption{Receive power versus number of antennas for a CW energy signal at 920 MHz and a transmit power of 79 mW per antenna.}
  \label{EH_power}
\end{figure}

\begin{figure}
   \centerline{\includegraphics[width=\columnwidth]{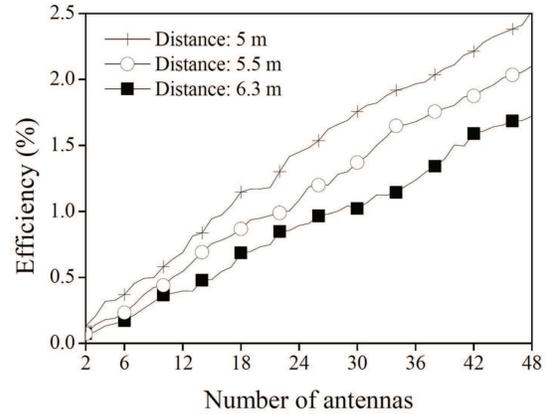}}
  \caption{End-to-end power transfer efficiency versus number of antennas for a CW energy signal at 920 MHz and a transmit power of 79 mW per antenna.}
  \label{EH_eff}
\end{figure}

Recently, a real-life multi-antenna wireless powered sensor network (WPSN) testbed has been reported in \cite{ChoiKim2017a}. A receive power-based channel estimation and energy beamforming algorithm \cite{ChoiKim2017b} for high RF power transfer efficiency and an adaptive duty cycle control algorithm for energy neutral operation at a sensor node have been implemented. Extensive experiments have been conducted to validate the feasibility of the multi-antenna WPSN and show the high performance of the proposed algorithms. To distribute RF power to multiple sensor nodes and keep them alive for perpetual operation, a real-life multi-node multi-antenna WPSN testbed has been implemented in \cite{ChoiKim2018a}. A joint beam-splitting and energy neutral control algorithm was designed by means of the Lyapunov optimization technique \cite{ChoiKim2016}. Experiments validate that the proposed algorithm can successfully keep all sensor nodes alive by optimally splitting energy beams towards multiple sensor nodes while maximizing the sum utility of the WPSN. To overcome the fundamental limit of RF power transfer and to enable deployment of battery-less sensors, a large-scale multi-antenna WPSN testbed was implemented at the Engineering Research Center of Sungkyunkwan University (ERC@SKKU). The following experiments were conducted: 1) beam-tracking, 2) beam-splitting, 3) energy neutral operation, 4) power transfer efficiency test (YouTube: https://youtu.be/qP9fZQX1sDk). In the end-to-end power transfer efficiency test, it was demonstrated that as the number of antennas grows, not only the total energy, but also the RF power transfer efficiency scales up, as shown in Figs. \ref{EH_power} and \ref{EH_eff}. This can be easily understood from Section \ref{antenna_model_subsection}, where the RF-to-DC conversion efficiency $e_3$ of state-of-the-art rectifiers was shown to increase as the average input power $P_{\rf}^r$ increases. The experimental results validate the benefit of multi-antenna beamforming and can be instrumental for the design and deployment of wireless-powered IoT sensors.

\par More recently, distributed RF power transfer system designs were reported in \cite{ChoiKim2018b} to overcome the low end-to-end power transfer efficiency. The corresponding experimental results confirm the theory \cite{Zeng:2017} and showed that using spatially distributed power beacons, each with a single antenna, can be advantageous in terms of the coverage probability over a single power beacon with many co-located antennas.


\subsection{Extension and Future Work}

Prototyping and experimentation of wireless power-based systems remains an important and much needed research area. Starting with WPT, some preliminary experimentation setups validating the benefits of designing signals (e.g. modulation, waveform, beamforming) specifically suited to maximize the harvested DC power are on-going as discussed in previous sections. Further experimental studies are nevertheless highly needed. A first research avenue that requires immediate attention is the validation of the experimental benefit of the new modulation formats discussed in Section \ref{NL_section}. Another promising area would be experimenting the performance benefits of combining all those different signals experimented separately in Sections \ref{SS_exp}, \ref{MS_exp} and \ref{MA_exp}. That would offer the joint benefit of the beamforming gain and the modulation/waveform gain. Hybrid design of the energy harvesters using a reconfigurable rectifier, which combines serial and parallel configurations of multiple energy harvesting circuits, may also be needed so as to further extend the range of applications. The serial configuration can improve $e_3$ for low RF input power, whereas the parallel one aims to increase $e_3$ for high RF input power \cite{Moon_Kim:2018}. Moreover, the prototyping and experimentation of multi-user WPT and validating the corresponding signal designs remains an important and uncharted research area.

\par When it comes to SWIPT, prototyping and experimentation is at an infancy stage with no experimental setup currently available to validate the R-E regions and the corresponding signal designs discussed in all previous sections. Circuit simulations have been used so far to validate some of the emerging SWIPT signal designs \cite{Clerckx:2016b,Clerckx:2017,Clerckx:2018b,Bayguzina:2018}. Some efforts are nevertheless on-going. Recently, a new transceiver architecture for dual mode SWIPT alternating between single-subband and multi-subband transmissions has been implemented, where the power management module monitors the harvested power and the power consumed by the information decoder with the aim of guaranteeing an energy neutral operation \cite{Park_Kim:2018,Park_Kim:2018a}. Experiments demonstrated that adaptive mode switching for dual mode operation improves the R-E tradeoff, compared to the conventional SWIPT. In \cite{Claessens:2017}, an integrated dual-purpose hardware to decode data and harvest energy is developed and the tradeoff between power and data reception quality is investigated. It is shown that the hardware can behave as a rectifier, depending on the information symbol rate and the cutoff frequency of the rectifier low-pass filter.

\section{Conclusions}\label{conclusions}
This article provides a tutorial overview on various energy harvester models and the corresponding signal and system designs for WIPT. The key conclusion of the paper is to highlight that WIPT signal and system designs crucially revolve around the underlying energy harvester model. Three different energy harvester models were presented, namely the conventional linear model, the diode nonlinear model and the saturation nonlinear model. Starting with single-user WIPT, we showed how the rate-energy region differs for the three different models and derived the corresponding transmitter and receiver architecture, waveform design, modulation, beamforming and input distribution optimizations, and resource allocation strategy. In particular, we showed that the fundamentals of PHY and MAC layer designs radically change in WIPT compared to existing communication systems because of the energy harvester nonlinearity. Moreover, some of those nonlinearities, such as the diode nonlinearity characterized by the diode nonlinear model, are actually beneficial to system performance and can be exploited to further enlarge the rate-energy region. We then turned our attention to multi-user WIPT and highlighted how the observations made for single-user extend to multi-user deployments. The validity of the different energy harvester models and the resulting signal designs were discussed and demonstrated using circuit simulations, prototyping, and experimentation. Throughout the manuscript, we also provided extensive discussions on promising research avenues. It is hoped that the techniques presented in this article will help inspiring future researches in this exciting new area and pave the way for designing and implementing efficient WIPT systems in the future.

\ifCLASSOPTIONcaptionsoff
  \newpage
\fi

\end{document}